\documentclass[preprint]{revtex4-1}

\usepackage[dvipsnames]{xcolor}

\usepackage{textcomp}

\usepackage{graphicx}
\usepackage{subfigure}
\usepackage{amsfonts}
\usepackage{amssymb}
\usepackage{latexsym}
\usepackage{natbib}
\usepackage{bm}
\usepackage{newlfont}
\usepackage{epsfig}
\usepackage{ctable}
\usepackage{amsmath}
\usepackage[american]{babel}
\usepackage{placeins}
\usepackage{cleveref}

\usepackage{algorithm}
\usepackage{algorithmic}

\crefname{section}{§}{§§}
\Crefname{section}{§}{§§}

\usepackage{bm}
\usepackage{sansmath}
\newcommand{\mathsfbi}[1]{{\sansmath{\bm{\mathsf{#1}}}}}

\usepackage{amsfonts}
\usepackage{amssymb}
\usepackage{amsmath}

\newcommand{\iddots}{%
  \mathinner{%
    \raisebox{0pt}{.}\kern1pt
    \raisebox{3pt}{.}\kern1pt
    \raisebox{6pt}{.}
  }%
}

\newcounter{alg}

\newenvironment{myalgorithm}[1]
{%
\refstepcounter{alg}
\begin{center}
\hrule
\vspace{0.5em}
\textbf{Algorithm \thealg: #1}
\vspace{0.5em}
\hrule
\vspace{0.5em}
\begin{minipage}{0.95\linewidth}
}
{%
\end{minipage}
\vspace{0.5em}
\hrule
\end{center}
}

\begin{document}

\title{\textcolor{black}{Projection method for mean resolvent analysis of periodic flows}}

\author{A.~Bongarzone}\email{alessandro.bongarzone@onera.fr} 
\author{C.~Content} 
\author{D.~Sipp} 
\author{C.~Leclercq}
\affiliation{ONERA, DAAA, Institut Polytechnique de Paris, 8 rue des Vertugadins, 92190 Meudon, France}

\begin{abstract}

\textcolor{black}{The mean resolvent operator predicts the mean linear response to forcing in the frequency domain and provides the optimal linear time-invariant approximation of statistically steady, time-varying flows \citep{leclercq2023mean}. We first leverage the harmonic resolvent framework \cite{wereley1990frequency,padovan2020analysis} in order to propose an algorithm for performing mean resolvent analysis of a periodic flow. Next, we propose an alternative approach which does not explicitly rely on the harmonic resolvent framework. The approach leverages the fact that the mean-flow resolvent approximates the mean resolvent operator, therefore the optimal forcing modes of the latter operator may be sought in a subspace spanned by optimal modes of the former. This projection approach does not require computing the adjoint dynamics about the attractor, which may be convenient for future extensions to more chaotic and turbulent flows. The present paper is however focused on periodic flows, where the convergence of the projection approach can be checked in comparison to the `ground truth' provided by the harmonic resolvent framework. This test is performed on a nearly incompressible axisymmetric laminar jet forced harmonically at the inlet. For the weakly unsteady case, the mean-flow resolvent captures the dominant receptivity peak but misses a secondary one present in the mean resolvent gain. For the strongly unsteady case, the mean-flow resolvent fails to predict the frequency of the vortex-pairing, while the mean resolvent correctly locates the corresponding gain peak. The projection method converges with a subspace dimension of 10 in the weakly unsteady case, while about 100 modes are required for accurate predictions in the strongly unsteady regime. Nonetheless, even a one-dimensional subspace correctly identifies the dominant receptivity peak.}

\end{abstract}

\maketitle

\begin{centering}\section{Introduction}\label{sec:INTRO}\end{centering}
\bigskip

The vast majority of industrial feedback control techniques require a linear time-invariant (LTI) model of the system's input-output (I/O) dynamics. Flow control makes no exception: most closed-loop control approaches tackling flow instabilities are also based on the LTI framework \citep{sipp2016linear}, although reinforcement learning is rapidly gaining ground \citep{fan2020reinforcement,vignon2023recent,xia2024active}. The transfer function from all inputs to all outputs (i.e. taking input and output matrices both equal to the identity) is called the resolvent operator. In the spatially-discrete framework, for any frequency, it corresponds to a large square matrix of dimension equal to the number of degrees of freedom of the discretised flow.\\

\begin{centering}\subsection{Resolvent analysis of unsteady flows: which operator ?}\end{centering}

Singular value decomposition (SVD) of the resolvent matrix provides, for any frequency, the optimal forcing modes leading to the greatest energy gain from input to output, as well as the corresponding optimal response modes. This procedure, called \textit{resolvent analysis} or \textit{input-output analysis} in the field of fluid mechanics, is key to understanding receptivity mechanisms in open flows: left/right singular modes provide information about optimal perturbation and response modes, and singular gains capture the optimal frequency range of non-modal energy amplification \citep{trefethen1993hydrodynamic,farrell1996generalized,schmid2002stability,jovanovic2005componentwise,sipp2013characterization,bugeat20193d,cook2024three}. Resolvent analysis also provides a low-rank representation of the operator which may be used for reduced-order modelling of the full linear system mapping all inputs to all outputs \citep{dergham2013stochastic}.\\
\indent \, LTI dynamics means linear dynamics about a fixed point, which is often called base flow in the fluid dynamics community. However, most flows of interest in aerodynamics are not evolving linearly in the vicinity of a steady base flow. Therefore, in order to apply resolvent analysis and any LTI feedback control tool to real-world flows, one needs to define meaningful ways to apply the LTI framework to unsteady flows evolving in a statistically steady regime. In this context, how do we even define a meaningful resolvent operator?\\

\begin{centering}\subsection{Mean-flow resolvent operator}\end{centering}

The common choice in the literature is to define the resolvent operator by linearizing the dynamics about the time-averaged mean flow \citep{mckeon2010critical,jeun2016input,beneddine2016conditions,lesshafft2019resolvent}, often adding a turbulent viscosity model to the molecular viscosity \citep{hwang2010amplification,hwang2010linear,morra2019relevance}, following a seminal idea for modal linear analysis of turbulent flows from Reynolds \& Hussain \citep{reynolds1972mechanics}. Neglecting turbulent viscosity creates an ill-posed resolvent operator. Indeed, the mean flow is not a fixed point of the Navier--Stokes equations and as a result, the poles of the mean-flow resolvent operator depend on the arbitrary choice of variables used for writing down the Navier--Stokes equations \citep{karban2020ambiguity}. However, the mean flow is a fixed point of the Reynolds-averaged Navier--Stokes (RANS) equations, so the resolvent operator is well-posed in this context. But applying the RANS framework to experimental or high-fidelity simulation data requires the determination of a turbulent viscosity field, which leads to new difficulties arising from the Boussinesq approximation. Indeed, determining a mean-flow-consistent eddy viscosity field from data is an overdetermined problem which can only be solved in a least-square sense, and the result may be locally negative or even unbounded \citep{rukes2016assessment,von2024role}. This problem can be solved by ad hoc clipping, regularisation or by making more hypotheses, i.e. assuming a turbulence closure model and fitting its coefficients from data \citep{rukes2016assessment,von2023self}. Despite its inherent limitations, this approach is, of course, invaluable from an engineering perspective. Most recent papers dealing with eddy viscosity in resolvent analysis assess the quality of their model by evaluating the ability of resolvent analysis to predict second-order statistics of unforced flow \citep{morra2019relevance}, and more specifically, to provide a good alignment between the leading resolvent mode and the leading spectral proper orthogonal decomposition (SPOD) mode \citep{pickering2021optimal,kuhn2022influence,symon2023use,mons2024data,thakor2024responses}. This alignment property is theoretically justified under stringent hypotheses \citep{beneddine2016conditions,towne2018spectral,lesshafft2019resolvent}. Following the seminal work of McKeon \& Sharma \citep{mckeon2010critical}, resolvent analysis has almost become synonymous with predicting unsteady coherent structures from a mean flow.\\
\indent \, However, we shall insist that the goal of the present paper is very distinct from that of predicting second-order statistics of unforced flows from their measured mean. Our end goal is to apply resolvent analysis to statistically-steady flows, in order to best characterise the linear response to \textit{exogenous} forcing, using a time-invariant approximation. None of the aforementioned studies addresses our specific problem, which calls for the introduction of the mean resolvent operator in the next paragraph.\\

\begin{centering}\subsection{Mean resolvent operator}\end{centering}

\begin{figure}
\centering
\includegraphics[width=0.8\textwidth]{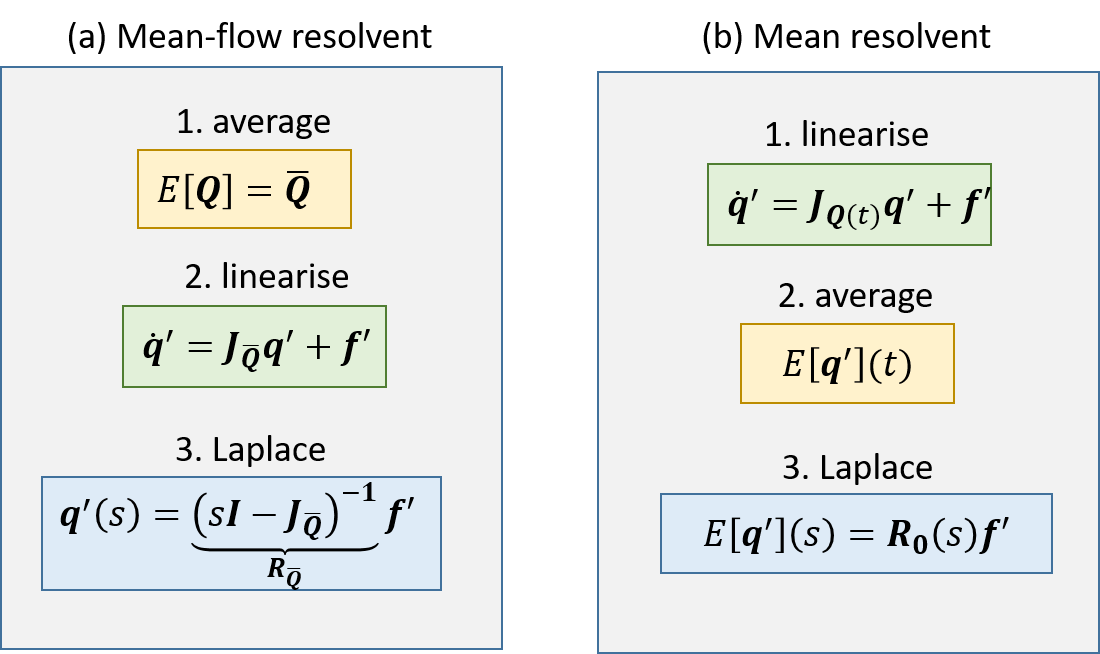}
\caption{Two ways of defining linear time-invariant input-output operators for statistically steady flows. The symbols $E[.]$ and $\overline{(.)}$ respectively denote ensemble and time averages. In the mean-flow resolvent approach (a), the dynamics is linearised about the time-averaged mean flow. In the mean resolvent approach, the dynamics is linearised about unsteady trajectories modelled as realisations of a stochastic process, and the resolvent operator predicts, in the frequency domain, the ensemble-averaged response to the deterministic forcing $\mathbf{f}'$. For periodic flows, the ensemble average is replaced by a phase average; stochasticity arises from the random choice of a phase to start forcing \citep{leclercq2023mean}.}
\label{fig:Fig0} 
\end{figure}

The mean resolvent operator predicts, in the frequency domain, the mean linear response to forcing about a flow evolving on an attractor \citep{leclercq2023mean}. The order of operations is reversed compared to the mean-flow resolvent, as illustrated in Fig.~\ref{fig:Fig0}. In the mean-flow resolvent framework, we first take an average, then linearise the dynamics about it, whereas in the mean resolvent framework, we linearise the dynamics about an ensemble of trajectories on the attractor, then average the distribution of linear responses to a given input. When switching the order of operations, we allow for the linear perturbation to interact not only with the mean flow, but also with the unsteady part of the underlying flow (which is also the case when injecting turbulent viscosity in the mean-flow resolvent framework). This operator is LTI if the underlying flow is statistically stationary. It is also optimal in a statistical sense, as it predicts the linear response with minimal error on (ensemble) average \citep{leclercq2023mean}. In the periodic case, ensemble averages amount to phase averages. In this case, it was also shown that the poles of the operator do not depend on an arbitrary choice of variables to write down the equations, unlike with the mean-flow resolvent. The mean resolvent framework is conceptually identical to linear response theory in the field of statistical physics (LRT; see Refs.~\onlinecite{marconi2008fluctuation,ruelle2009review} for reviews), where mean transfer functions are called admittances \citep{marconi2008fluctuation} or susceptibilities \citep{ruelle2009review}. Although the mean-flow resolvent approximates the mean resolvent operator in the weakly-unsteady limit for periodic flows \citep{leclercq2023mean}, in general, they truly are different objects. Russo \& Luchini \citep{russo2016linear} indirectly showed that it is impossible to derive a physically-meaningful eddy viscosity that would make the mean resolvent and the mean-flow resolvent equal, in the case of turbulent channel flow. Indeed, the authors showed that making the mean linear response to steady forcing equal to the linear response about the mean flow requires locally negative and infinite eddy viscosities, underlining again a limitation of the Boussinesq approximation.\\

\textcolor{black}{\begin{centering}\subsection{Mean resolvent analysis beyond the harmonic resolvent framework}\end{centering}}

Following the seminal work by Hussain \& Reynolds \citep{hussain1970mechanics}, single-input single-output mean transfer functions from localized actuators to localized sensors have been measured or computed on countless occasions, mostly for closed-loop control purposes \citep{mongeau1998active,kestens1998active,cattafesta1999development,kook2002active,kegerise2002adaptive,rathnasingham2003active,cabell2006experimental,dahan2012feedback,maia2021real,leclercq2023mean,audiffred2024reactive,jussiau2024data}. Multiple-input multiple-output (MIMO) cases have also been considered, but only in the time-domain: Luchini \& Quadrio \citep{luchini2006phase} computed the mean linear impulse response of turbulent channel flow subjected to boundary forcing (this was later used for turbulent drag reduction through optimal control by Martinelli, Quadrio \& Luchini \citep{martinelli2009turbulent}). Carini \& Quadrio \citep{carini2010direct} and Matsumoto \textit{et al.} \citep{matsumoto2021correlation} computed mean impulse responses in homogeneous isotropic turbulence. \cite{unnikrishnan2016high} and \cite{adler2018dynamic} computed the mean linear response to stochastic forcing of a supersonic jet and a shock-wave boundary-layer interaction.\\
\indent \, In the present paper, we are interested in obtaining, in the frequency domain, a low-rank representation of the operator mapping any (volumic) forcing to its corresponding (volumic) mean response, in a spatially inhomogeneous case. In other words, we seek efficient methods for performing the SVD of the high-dimensional mean resolvent matrix. SVD solves an optimisation problem, which requires multiple applications of the resolvent operator and its adjoint. \textcolor{black}{Matrix-based \citep{sipp2013characterization,ribeiro2020randomized,house2022sketch} and matrix-free \citep{monokrousos2010global,martini2021efficient,farghadan2025scalable} implementations are available for SVD of the mean-flow resolvent operator. Similarly, matrix-based \citep{padovan2020analysis,padovan2022analysis} and matrix-free \citep{farghadan2024efficient} implementations are available for SVD of the harmonic resolvent operator. Given the strong connection between the harmonic resolvent and the mean resolvent operators in the periodic case \cite{leclercq2023mean}, it is straightforward to adapt these methods for SVD of the mean resolvent operator, as we shall see in \S \ref{sec:Sec1}. However, the focus of the paper is elsewhere. Indeed, while the harmonic resolvent operator is specific to periodic flows, the mean resolvent operator generalizes to all statistically steady flows. Our goal is therefore to propose a method which may later be generalized to stochastic, chaotic and turbulent flows. The issue with the matrix-based approach from harmonic resolvent analysis is that it relies on the harmonic balance framework, which is indeed restricted to periodic dynamics. The issue with the matrix-free approach is that it relies on time-integration of the adjoint equations about \textit{unsteady} trajectories, which may not be easily generalized to chaotic dynamics \cite{blonigan2018multiple}.\\
\indent \, In the present paper, we therefore propose a \textit{projection method} which does not suffer from these limitations. This alternative approach leverages the fact that the mean-flow resolvent approximates the mean resolvent \cite{leclercq2023mean}. Accordingly, we will make the assumption that the optimal forcing modes of the mean resolvent belong to a small subspace of optimal forcing modes of the mean-flow resolvent. Under this assumption, we will be able to perform mean resolvent analysis \textit{without} evaluating the action of the adjoint linearized operator about the unsteady flow, whether the implementation is matrix-based or matrix-free. Adjoint methods are still required, but only in the pre-processing step of computing optimal forcing modes of the resolvent operator about the mean flow, which poses no major difficulty, even for chaotic flows. Even though our long term goal is to extend the methodology to more complex flows, we solely investigate periodic flows in the present paper. The goal is to study the convergence of the projection approach with respect to the subspace dimension, against 'ground truth' results available from the harmonic resolvent framework.}

\begin{centering}\subsection{Organisation of the manuscript}\end{centering}

In \S \ref{sec:Sec1}, we provide theoretical background on the mean-flow resolvent and mean resolvent operators in the context of periodic flows. \textcolor{black}{We describe adjoint-based approaches for mean-flow and mean resolvent analysis, then describe the projection method which does not require evaluating the adjoint dynamics about unsteady trajectories (only the adjoint about the mean flow is needed). In all cases, we present matrix-based and matrix-free implementations.} In~\S\ref{sec:Sec3}, we apply the adjoint-based approach to a (nearly-)incompressible laminar axisymmetric jet periodically forced at the inlet \citep{padovan2022analysis}, and compare the results with mean-flow resolvent analysis. In~\S\ref{sec:Sec4sub1}, we implement the projection-based approach and study its convergence against the results from~\S\ref{sec:Sec3}. Conclusions and outlook are provided in \S\ref{sec:CONCL}.\\


\begin{centering}\section{Linear input-output analysis of periodic flow}\label{sec:Sec1}\end{centering} 
 
We consider a nonlinear dynamical system under the influence of a $T_0$-periodic forcing $\mathbf{g}$:
\begin{equation}
\label{eq:wqBF}
\frac{\mathrm{d}\mathbf{q}}{\mathrm{d}t}=\boldsymbol{\mathrm{r}}\left(\mathbf{q}\left(t\right)\right)+\mathbf{g}\left(t\right).
\end{equation}
In the present paper, this dynamical system represents the forced compressible Navier--Stokes equations. Whether $\mathbf{g}(t)$ is zero or not, we assume that the system admits a $T_0$-periodic solution denoted $\mathbf{Q}(t)$.\\
\indent \, We now consider the linear response $\mathbf{q}'$ to infinitesimal forcing $\mathbf{g}'$ about the basic flow $\mathbf{Q}$. The forcing starts at an arbitrary time $t=t_0$, which is associated with a particular phase $0\leq \phi<2\pi$ of the basic flow such that
\begin{equation}
    \omega_0t_0= \phi\,\mathrm{mod}2\pi,
\end{equation}
with $\omega_0=2\pi/T_0$. The time variable will now be shifted $t:=t-t_0$ so that the origin of time always matches with the start of the forcing, regardless of the phase $\phi$, which becomes a parameter; in practice, for us, a random variable uniformly distributed in $\left[0,2\pi\right)$. We shall explicitly keep track of the phase-dependence in the rest of this paper; for instance, the basic flow and perturbations will be denoted $\mathbf{Q}(t;\phi)$, $\mathbf{q}'(t;\phi)$ to highlight this parametric dependence, which arises from the arbitrary choice of $t_0$. The Jacobian $\mathsfbi{J}_{\boldsymbol{\mathrm{Q}}}\left(t;\phi\right)=\left.\mathrm{d}_{\boldsymbol{\mathrm{q}}} \boldsymbol{\mathrm{r}}\right|_{\boldsymbol{\mathrm{Q}}\left(t;\phi\right)}$ about the periodic basic flow is also $T_0$-periodic and parametrized by $\phi$. Even though the resolvent operator is, strictly speaking, from all $\mathrm{N}$ inputs to all $\mathrm{N}$ outputs, with $\mathrm{N}$ being the dimension of the state-vector $\mathbf{q}$, we will restrict the input and output variables to respectively $\mathrm{M}$ and $\mathrm{K}$ physically-relevant degrees of freedom, with $\mathrm{M}<\mathrm{N}$ and $\mathrm{K}<\mathrm{N}$. This is done by introducing an input matrix $\mathsfbi{B}\in\mathbb{R}^{\mathrm{N}\times \mathrm{M}}$ and an output matrix $\mathsfbi{C}\in\mathsfbi{R}^{\mathrm{K}\times \mathrm{N}}$. Each column (respectively line) of $\mathsfbi{B}$ (respectively $\mathsfbi{C}$) is zero except for a single component equal to one, selecting a different degree of freedom for each column (respectively row); therefore $\mathsfbi{B}$ and $\mathsfbi{C}$ will also be referred to as prolongation and restriction matrices. Restricted input $\mathbf{f}'$ and output $\mathbf{y}'$ variables satisfy $\mathbf{g}'=\mathsfbi{B}\mathbf{f}'$ and $\mathbf{y}'=\mathsfbi{C}\mathbf{q}'$, and the linear input-output dynamics is governed by the following equations:
\begin{align}
\frac{\mathrm{d}\boldsymbol{\mathrm{q}}'}{\mathrm{d}t}\left(t;\phi\right)&=\mathsfbi{J}_{\boldsymbol{\mathrm{Q}}}\left(t;\phi\right)\boldsymbol{\mathrm{q}}'\left(t;\phi\right)+\mathsfbi{B}\boldsymbol{\mathrm{f}}'\left(t\right),\label{eq:eq3MR}\\
\boldsymbol{\mathrm{y}}'\left(t;\phi\right)&=\mathsfbi{C}\boldsymbol{\mathrm{q}}'\left(t;\phi\right).
\label{eq:eq3bisMR}
\end{align}

\indent \textcolor{black}{Although a Lyapunov--Floquet transformation may be employed to recast equations (\ref{eq:eq3MR}-\ref{eq:eq3bisMR}) into a convenient time-invariant (LTI) form, so as to perform the input-output (I/O) analysis directly in these Floquet coordinates \citep{lin2023flow}, before mapping back to physical space, the actual dynamics in the latter space is not time-invariant because the transformation is time-dependent; more precisely, the dynamics described by (\ref{eq:eq3MR}-\ref{eq:eq3bisMR}) where the Jacobian is periodic is called linear time-periodic (LTP).} Also, the response of the time-varying system to forcing depends on the initial time $t_0$, through the phase $\phi$ (we stress that $\mathbf{f}'$ itself does not depend on $\phi$). In the following, we will present the two linear time-invariant (LTI) approximations \textcolor{black}{(in physical space)} of the I/O dynamics graphically introduced in Fig.~\ref{fig:Fig0}.\\



\begin{centering}\subsection{Mean-flow resolvent operator}\label{subsec:Sec1sub1}\end{centering}

As depicted in Fig.~\ref{fig:Fig0}, a straightforward LTI approximation of the LTP dynamics is obtained by discarding the phase-dependent base flow harmonics and only retaining the time-averaged flow field $\overline{\boldsymbol{\mathrm{Q}}}=\left(1/T_0\right)\int_{0}^{T_0}\boldsymbol{\mathrm{Q}}\left(t\right)\,\mathrm{d}t$; the instantaneous Jacobian in~\eqref{eq:eq3MR} is then replaced by the Jacobian about the mean flow $\mathsfbi{J}_{\overline{\boldsymbol{\mathrm{Q}}}}$.\\
\indent \, Assuming $\mathbf{y}'(t=0)=0$, the linear response to forcing about the mean flow is provided in the frequency domain by 
\begin{equation}
    \mathbf{y}'(s)=\mathsfbi{R}_{\overline{\boldsymbol{\mathrm{Q}}}}(s)\mathbf{f}'(s),
\end{equation}
where $s$ is the Laplace variable and
\begin{equation}
\label{eq:eq15quadris}
\mathsfbi{R}_{\overline{\boldsymbol{\mathrm{Q}}}}\left(s\right)=\mathsfbi{C}\left(s\mathsfbi{I}-\mathsfbi{J}_{\overline{\boldsymbol{\mathrm{Q}}}}\right)^{-1}\mathsfbi{B}
\end{equation}
is the resolvent operator about the mean flow ($\mathsfbi{I}$ denotes the identity, but could be replaced in practice by a positive mass matrix depending on the spatial discretisation scheme).\\

\begin{centering}\subsection{Mean resolvent operator}\label{subsec:Sec1sub0}\end{centering}

Using Floquet theory, Leclercq \& Sipp \citep{leclercq2023mean} showed that, in the frequency domain, the linear response to forcing of the LTP system (\ref{eq:eq3MR}-\ref{eq:eq3bisMR}) to an arbitrary forcing $\mathbf{f}'(t)$ is given by
\begin{equation}
    \mathbf{y}'(s;\phi)=\sum_n \mathrm{e}^{\mathrm{i}n\phi}\mathsfbi{R}_n(s)\mathbf{f}'(s-\mathrm{i}n\omega_0),\label{eq:LTP}
\end{equation}
where
\begin{equation}
\label{eq:BLOCKn}
\mathsfbi{R}_n\left(s\right)=\mathsfbi{C}\left(\sum_{j}\sum_{k=1}^{\mathrm{N}}\frac{\hat{\boldsymbol{\mathrm{v}}}_j^k\left(\hat{\boldsymbol{\mathrm{w}}}_{j-n}^k\right)^H}{s-\left(\lambda_k+\text{i}j\omega_0\right)}\right)\mathsfbi{B}.
\end{equation}
The vector $\hat{\boldsymbol{\mathrm{w}}}_{j}^k$ is the $j$th Fourier component of the $k$th adjoint Floquet mode characterising receptivity to the corresponding Fourier component $\hat{\boldsymbol{\mathrm{v}}}_j^k$ of the associated direct Floquet mode. The poles $\lambda_k$ of $\mathsfbi{R}_n$ are exactly the Floquet exponents of the LTP system. Averaging \eqref{eq:LTP} with respect to $\phi$ yields 
\begin{equation}
    \langle\mathbf{y}'(s)\rangle_{\phi}=\mathsfbi{R}_0(s)\mathbf{f}'(s)\label{eq:avg},
\end{equation} 
so $\mathsfbi{R}_0$ is, by definition, the mean resolvent operator. Expression (\ref{eq:BLOCKn}) is arguably not the most convenient way to compute the effect of the mean resolvent or its adjoint on an input vector. In the next two subsections, we will propose two practical ways to perform these operations: a matrix-based approach using linear algebra (\S \ref{subsec:freqdomain}) and a matrix-free version using time-stepping (\S \ref{subsec:timedomain}).\\
\indent \, The connection between mean-flow resolvent and mean resolvent operators can be made explicit using a Neumann series \citep{leclercq2023mean}. To write down the expansion, we first need to introduce the Fourier series of the periodic Jacobian
\begin{equation}
\label{eq:eq6}
\mathsfbi{J}_{\boldsymbol{\mathrm{Q}}}\left(t;\phi\right)=\overline{\mathsfbi{J}}+\sum_{n\ne 0} \hat{\mathsfbi{J}}_n \mathrm{e}^{\mathrm{i}n\phi}  \mathrm{e}^{\mathrm{i}n\omega_0 t}
\end{equation}
and the notation $\mathsfbi{R}_{\overline{\mathsfbi{J}}}=(s\mathsfbi{I}-\overline{\mathsfbi{J}})^{-1}$. The Neumann series
\begin{align}
    \mathsfbi{R}_0(s)&=\mathsfbi{R}_{\overline{\mathsfbi{J}}}(s)\label{eq:Neumann}\\
    &+\sum_j \hat{\mathsfbi{J}}_j^*\mathsfbi{R}_{\overline{\mathsfbi{J}}}(s+\mathrm{i}j\omega_0)\hat{\mathsfbi{J}}_j\mathsfbi{R}_{\overline{\mathsfbi{J}}}(s)\nonumber\\
    &+\sum_{j,k}\hat{\mathsfbi{J}}_j^*\mathsfbi{R}_{\overline{\mathsfbi{J}}}(s+\mathrm{i}j\omega_0)\hat{\mathsfbi{J}}_{j-k}\mathsfbi{R}_{\overline{\mathsfbi{J}}}(s+\mathrm{i}k\omega_0)\hat{\mathsfbi{J}}_k\mathsfbi{R}_{\overline{\mathsfbi{J}}}(s)\nonumber\\
    &+\dots\nonumber,
\end{align}
converges in some right-half plane \citep{leclercq2023mean}, where $()^*$ stands for the complex conjugate (not the Hermitian transpose). When nonlinearities are quadratic, as in the incompressible Navier--Stokes equations, $\overline{\mathsfbi{J}}=\mathsfbi{J}_{\overline{\mathsfbi{Q}}}$, therefore $\mathsfbi{R}_{\overline{\mathsfbi{J}}}=\mathsfbi{R}_{\overline{\mathbf{Q}}}$. In that case, the series expansion shows how the mean-flow resolvent approximates the mean resolvent in the weakly unsteady limit where the unsteady part of the Jacobian may be neglected. In general though, whether $\mathsfbi{R}_{\overline{\mathsfbi{J}}}=\mathsfbi{R}_{\overline{\mathbf{Q}}}$ or not, the two operators $\mathsfbi{R}_0$ and $\mathsfbi{R}_{\overline{\mathbf{Q}}}$ are different because the mean resolvent accounts for interactions of the linear perturbation with the unsteady part of the base flow, not just the mean flow.\\

\begin{centering}\subsubsection{Matrix-based implementation: prolongation/restriction on the harmonic resolvent}\label{subsec:freqdomain}\end{centering}

The matrix-based approach to evaluating the mean resolvent is based on the harmonic resolvent matrix, which was initially introduced by Wereley \& Hall \citep{wereley1990frequency,wereley1991linear} and brought to the fore in the fluid dynamics community by Padovan, Otto \& Rowley \citep{padovan2020analysis}. Refs.~\onlinecite{wereley1990frequency,wereley1991linear} consider LTP systems with basic forcing and responses in the form of exponentially-modulated periodic (EMP) signals, i.e. 
\begin{equation}
 \mathbf{f}'=\mathrm{e}^{st}\sum_n\hat{\mathbf{f}}_n \mathrm{e}^{\mathrm{i}n\omega_0 t}\quad\text{and}\quad\mathbf{y}'=\mathrm{e}^{st}\sum_n\hat{\mathbf{y}}_n(s;\phi) \mathrm{e}^{\mathrm{i}n\omega_0 t}.\label{eq:EMP} 
\end{equation}
Plugging (\ref{eq:EMP}) and (\ref{eq:eq6}) into the LTP system (\ref{eq:eq3MR}-\ref{eq:eq3bisMR}) and using a harmonic balance procedure \citep{khalil2002nonlinear}, one finds an infinite-dimensional linear mapping relating the frequency coefficients of the input to the frequency coefficients of the output
\begin{equation}
\label{eq:eq13}
\hat{\boldsymbol{\mathsfbi{\mathcal{Y}}}}\left(s;\phi\right)=
\boldsymbol{\mathsfbi{\mathcal{R}}}(s;\phi)
\hat{\boldsymbol{\mathsfbi{\mathcal{F}}}}, \,\,\,\, \,\,\,\, \,\,\,\,
\hat{\boldsymbol{\mathsfbi{\mathcal{Y}}}}\left(s;\phi\right)=
\left(
\begin{matrix}
\vdots\\
\hat{\boldsymbol{\mathrm{y}}}_{-1}(s;\phi)\\
\hat{\boldsymbol{\mathrm{y}}}_0(s;\phi)\\
\hat{\boldsymbol{\mathrm{y}}}_{+1}(s;\phi)\\
\vdots
\end{matrix}\right), \,\,\,\, \,\,\,\, \,\,\,\,
\hat{\boldsymbol{\mathsfbi{\mathcal{F}}}}=
\left(
\begin{matrix}
\vdots\\
\hat{\boldsymbol{\mathrm{f}}}_{-1}\\
\hat{\boldsymbol{\mathrm{f}}}_0\\
\hat{\boldsymbol{\mathrm{f}}}_{+1}\\
\vdots
\end{matrix}\right),
\end{equation}
where
\begin{align}
\boldsymbol{\mathsfbi{\mathcal{R}}}(s;\phi)&=\boldsymbol{\mathsfbi{\mathcal{C}}}(s\boldsymbol{\mathcal{I}}-\boldsymbol{\mathcal{H}}(\phi))^{-1}\boldsymbol{\mathsfbi{\mathcal{B}}},\label{eq:harmres}\\
\boldsymbol{\mathcal{C}}&=\mathrm{blkdiag}\left(\dots,\mathsfbi{C},\mathsfbi{C},\mathsfbi{C},\dots\right),\\
\boldsymbol{\mathcal{B}}&=\mathrm{blkdiag}\left(\dots,\mathsfbi{B},\mathsfbi{B},\mathsfbi{B},\dots\right),\\
\boldsymbol{\mathcal{I}}&=\mathrm{blkdiag}\left(\dots,\mathsfbi{I},\mathsfbi{I},\mathsfbi{I},\dots\right),
\end{align}
\begin{align}
\boldsymbol{\mathcal{H}}(\phi)&=\left( 
\begin{matrix}
\ddots & \vdots & \vdots & \vdots & \iddots \\
\hdots & \overline{\mathsfbi{J}}-\mathrm{i}\omega_0\mathsfbi{I} & \hat{\mathsfbi{J}}_{-1}\mathrm{e}^{-\mathrm{i}\phi} & \hat{\mathsfbi{J}}_{-2}\mathrm{e}^{-\mathrm{i}2\phi} & \hdots\\
\hdots & \hat{\mathsfbi{J}}_{+1}\mathrm{e}^{+\mathrm{i}\phi} &\overline{\mathsfbi{J}} & \hat{\mathsfbi{J}}_{-1}\mathrm{e}^{-\mathrm{i}\phi} & \hdots\\
\hdots & \hat{\mathsfbi{J}}_{+2}\mathrm{e}^{+\mathrm{i}2\phi} & \hat{\mathsfbi{J}}_{+1}\mathrm{e}^{+\mathrm{i}\phi} & \overline{\mathsfbi{J}}+\mathrm{i}\omega_0\mathsfbi{I} & \hdots\\
\iddots & \vdots & \vdots & \vdots & \ddots \\
\end{matrix}\label{eq:Hill}
\right).
\end{align}
The operator $\boldsymbol{\mathcal{R}}$ is the harmonic transfer function, and the infinite matrix $\boldsymbol{\mathcal{H}}$ is called the Hill matrix \citep{lazarus2010harmonic,franceschini2022identification}. Using \eqref{eq:LTP} it may also be shown that
\begin{equation}
\boldsymbol{\mathcal{R}}(s;\phi)=\left(
\begin{matrix}
\ddots & \vdots & \vdots & \vdots & \iddots \\
\hdots & \mathsfbi{R}_0\left(s-\mathrm{i}\omega_0\right) & \mathsfbi{R}_{-1}\left(s-\mathrm{i}\omega_0\right)\mathrm{e}^{-\mathrm{i}\phi} & \mathsfbi{R}_{-2}\left(s-\mathrm{i}\omega_0\right)\mathrm{e}^{-\mathrm{i}2\phi} & \hdots\\
\hdots & \mathsfbi{R}_{+1}\left(s\right)\mathrm{e}^{+\mathrm{i}\phi} &\mathsfbi{R}_0\left(s\right) & \mathsfbi{R}_{-1}\left(s\right)\mathrm{e}^{-\mathrm{i}\phi} & \hdots\\
\hdots & \mathsfbi{R}_{+2}\left(s+\mathrm{i}\omega_0\right)\mathrm{e}^{+\mathrm{i}2\phi} & \mathsfbi{R}_{+1}\left(s+\mathrm{i}\omega_0\right)\mathrm{e}^{+\mathrm{i}\phi} & \mathsfbi{R}_0\left(s+\mathrm{i}\omega_0\right) & \hdots\\
\iddots & \vdots & \vdots & \vdots & \ddots \\
\end{matrix}\right),\label{eq:alsobeshown},
\end{equation}
therefore, the mean resolvent operator appears along the diagonal of the harmonic resolvent and may be expressed as 
\begin{equation}
\boldsymbol{\mathsfbi{R}}_0(s)=\boldsymbol{\mathsfbi{\mathcal{P}}}_0^T\boldsymbol{\mathcal{R}}(s;\phi)\boldsymbol{\mathsfbi{\mathcal{P}}}_0,\label{eq:meanres}
\end{equation}
where the matrix
\begin{equation}
\boldsymbol{\mathcal{P}}^T_0=\left(\dots,\mathsfbi{0},\mathsfbi{I},\mathsfbi{0},\dots\right)
\end{equation}
is used to extract the `00' block of $\boldsymbol{\mathcal{R}}$. The method relies on the explicit computation of the Fourier expansion of the periodic Jacobian, in order to form the Hill matrix. Numerical methods to efficiently perform the matrix vector products $\mathbf{x}=\mathsfbi{R}_0 \mathbf{b}$ and $\mathbf{x}=\mathsfbi{R}_0^H \mathbf{b}$ based on the above expressions are left to section \S\ref{subsec:linalg}.\\

\begin{centering}\subsubsection{Matrix-free implementation: harmonic average of linear response to harmonic forcing}\label{subsec:timedomain} \end{centering}

It is also possible to evaluate the effect of the mean resolvent without forming the harmonic resolvent matrix. Consider, for that, the case of linear harmonic forcing $\mathbf{f}'(t\geq 0)=\hat{\mathbf{f}}_0\mathrm{e}^{\mathrm{i}\omega t}+\mathrm{c.c.}$ about a periodic base flow which is not self-sustained (i.e. $\mathbf{g}\neq 0$), such that all Floquet exponents are damped \citep{leclercq2023mean}. Then the permanent EMP response $\mathbf{y}'(t;\phi)=\mathrm{e}^{\mathrm{i}\omega t}\sum_n \hat{\mathbf{y}}_n(\phi) \mathrm{e}^{\mathrm{i}n\omega_0 t}$ to that specific EMP input is obtained by taking $s=\mathrm{i}\omega$ and $\hat{\boldsymbol{\mathsfbi{\mathcal{F}}}}=(\dots,0,\hat{\mathbf{f}}^T_0,0,\dots)^T$ in (\ref{eq:eq13}-\ref{eq:alsobeshown}):
\begin{equation}
\hat{\mathbf{y}}_n(\phi)=\mathsfbi{R}_n(\mathrm{i}(\omega+n\omega_0))\mathrm{e}^{\mathrm{i}n\phi}\hat{\mathbf{f}}_0,\quad\forall n\in\mathbb{Z}.
\end{equation}
In particular, for $n=0$, we have
\begin{equation}
\hat{\mathbf{y}}_0=\mathsfbi{R}_0(\mathrm{i}\omega)\hat{\mathbf{f}}_0,\label{eq:EMPresp},
\end{equation}
so that extracting the component $\hat{\mathbf{y}}_0$ at the forcing frequency $\omega$ from the linear response amounts to applying the mean resolvent to $\hat{\mathbf{f}}_0$. Similarly, one may evaluate the effect of the adjoint mean resolvent operator by extracting the component at $\omega$ from the linear response of the adjoint linearised equations to harmonic forcing.\\
\indent \, For a broadband input $\mathbf{f}'\left(t\right)$, computing the phase-averaged response to arbitrary forcing requires ensemble averaging over multiple realisations for various phases at $t_0$. However, in the case of harmonic input, a much cheaper option simply consists in evaluating the harmonic average
\begin{equation}
    \hat{\mathbf{y}}_0=\lim_{T\to \infty}\dfrac{1}{T}\int_0^T \mathbf{y}'(t;\phi)\mathrm{e}^{-\mathrm{i}\omega t}\,\mathrm{d}t.\label{eq:harmavg}
\end{equation}
from a \textit{single} realization of the forcing. The term dynamic linearity, coined by Dahan, Morgans \& Lardeau \citep{dahan2012feedback}, was used in Ref.~\onlinecite{leclercq2023mean} to describe this convenient property of harmonic forcings in the LTP context, i.e. no need for ensemble averaging for that class on input signals on that class of systems. The harmonic average may be computed on-the-fly to avoid storing the timeseries $\mathbf{y}'(t)$.\\

\begin{centering}\subsection{Adjoint-based method for resolvent analysis}\end{centering}

Whether $\mathsfbi{R}=\mathsfbi{R}_{\overline{\mathbf{Q}}}(\mathrm{i}\omega)$ or $\mathsfbi{R}=\mathsfbi{R}_0(\mathrm{i}\omega)$, resolvent analysis consists in finding the non-trivial harmonic forcing $\mathbf{f}=\hat{\mathbf{f}}_0\mathrm{e}^{\mathrm{i}\omega t}+\mathrm{c.c.}$ maximizing the energy gain $G=\|\mathsfbi{R}\hat{\mathbf{f}}_0\|_{\mathsfbi{M}_y}^2/\|\hat{\mathbf{f}}_0\|_{\mathsfbi{M}_f}^2$, where $\mathsfbi{M}_f$ and $\mathsfbi{M}_y$ are respectively positive-definite and positive semi-definite inner product matrices. The optimal forcing is the eigenvector of the following Hermitian generalised eigenvalue problem
\begin{equation}
\mathsfbi{R}^H\mathsfbi{M}_y\mathsfbi{R}\hat{\boldsymbol{\psi}}_j=\lambda_j^2\mathsfbi{M}_f\hat{\boldsymbol{\psi}}_j\label{eq:ghevp}
\end{equation}
associated with the greatest eigenvalue $\lambda_1^2(\geq\lambda_2^2\geq \dots \geq 0)$, which is equal to the optimal energy gain. Eigenpairs for $j\geq 2$ correspond to suboptimal gains and forcings. The forcing modes are normalised with respect to $\mathsfbi{M}_f$ and form an $\mathsfbi{M}_f$-orthogonal family. The corresponding (sub)-optimal response modes defined by $\hat{\boldsymbol{\phi}}_j=\mathsfbi{R}\hat{\boldsymbol{\psi}}_j/\|\mathsfbi{R}\hat{\boldsymbol{\psi}}_j\|_{\mathsfbi{M}_y}$ form a $\mathbf{M}_y$-orthogonal family. In the rest of this paper, we will denote with a subscript $(.)_{MF}$ results belonging to mean-flow resolvent analysis, and with a subscript $(.)_{MR}$ results belonging to mean resolvent analysis.\\
\indent \, Finding the eigenvalues of greatest magnitude of the eigenvalue problem (\ref{eq:ghevp}) requires the application of some variant of the power method. This method, and Krylov-based generalisations of it such as the Arnoldi method, require efficient evaluation of the matrix-vector products $\mathbf{x}=\mathsfbi{R}^H\mathsfbi{M}_y\mathsfbi{R}\mathbf{b}$ (and $\mathbf{z}=\mathsfbi{M}_f \mathbf{c}$). There are two ways to proceed: either using sparse linear algebra or using adjoint-looping with a time-stepper. These two approaches are respectively presented in \S \ref{subsec:linalg} and \S \ref{subsec:timestep} below. The optimal gains and modes that we solve for are related to the singular value decomposition of $\mathsfbi{M}_y^{1/2}\mathsfbi{R}\mathsfbi{M}_f^{-1/2}$. More precisely, if we denote $(\sigma_j,\mathbf{r}_j,\mathbf{l}_j)$ the $j^\textsuperscript{th}$ singular value, right and left singular vector, we have: $\lambda_j^2=\sigma_j^2$, $\mathbf{r}_j=\mathsfbi{M}_f^{1/2}\hat{\boldsymbol{\psi}}_j$ and $\mathbf{l}_j=\mathsfbi{M}_y^{1/2}\hat{\boldsymbol{\phi}}_j$.\\

\begin{centering}\subsubsection{Matrix-based implementation: Arnoldi algorithm with sparse linear solvers}\label{subsec:linalg}\end{centering}

In the linear algebra paradigm, multiple eigenvalues $\lambda_j^2$ may be converged at once using the Arnoldi algorithm implemented in SLEPc. For the mean-flow resolvent case, this implies two linear system solves: $(\mathrm{i}\omega\mathsfbi{I}-\mathsfbi{J}_{\overline{\mathbf{Q}}})\mathbf{y}=\mathbf{b}$ and $(\mathrm{i}\omega\mathsfbi{I}-\mathsfbi{J}_{\overline{\mathbf{Q}}})^H\mathbf{x}=\mathbf{y}$. This is efficiently done using a LU decomposition of $(\mathrm{i}\omega \mathsfbi{I}-\mathsfbi{J}_{\overline{\mathbf{Q}}})$ using the sparse solver MUMPs \citep{amestoy2000mumps}, implemented in PETSc \citep{balay2019petsc}. \textcolor{black}{It is worth mentioning that alternative approaches also exist, based on randomized singular value decomposition (RSVD-LU) techniques \citep{ribeiro2020randomized,house2022sketch}, which approximate the dominant resolvent modes and gains by projecting the operator onto a low-dimensional random subspace, thereby reducing the computational cost associated with large-scale SVD computations.}\\
\indent \, For the mean resolvent case, one needs to perform the matrix-vector product $\mathbf{x}=$ $\boldsymbol{\mathcal{P}}_0^T\boldsymbol{\mathcal{B}}^T(\mathrm{i}\omega \boldsymbol{\mathcal{I}}-\boldsymbol{\mathcal{H}})^{-H}\boldsymbol{\mathcal{C}}^T\boldsymbol{\mathcal{P}}_0\mathsfbi{M}_y\boldsymbol{\mathcal{P}}_0^T\boldsymbol{\mathcal{C}}(\mathrm{i}\omega \boldsymbol{\mathcal{I}}-\boldsymbol{\mathcal{H}})^{-1}\boldsymbol{\mathcal{B}}\boldsymbol{\mathcal{P}}_0\mathbf{b}$, which implies solving block linear systems of the form $(\mathrm{i}\omega \boldsymbol{\mathcal{I}}-\boldsymbol{\mathcal{H}})\mathbf{r}=\mathbf{z}$ and $(\mathrm{i}\omega \boldsymbol{\mathcal{I}}-\boldsymbol{\mathcal{H}})^H\mathbf{u}=\mathbf{w}$. These block linear systems are solved in an iterative fashion, using the Generalised Minimal Residual (GMRES) algorithm \citep{saad1986gmres}. In this study, the code is made parallel and distributed using PETSc, with each processor core handling one harmonic of the response, i.e. a line of blocks. In order to accelerate convergence for poorly conditioned problems, a simple block-Jacobi preconditioner is used to approximate the inverse of the block matrix $\mathrm{i}\omega \boldsymbol{\mathcal{I}}-\boldsymbol{\mathcal{H}}$, as suggested in Refs.~\onlinecite{rigas2021nonlinear,poulain2024adjoint}. This amounts to inverting the blocks $[\mathrm{i}\left(\omega+n\omega_0\right)\mathsfbi{I}-\overline{\mathsfbi{J}}]$ along the diagonal.\\
\indent \, This inversion is handled by each processor using a sparse direct LU method (MUMPS) and therefore only requires storage of the sparse block on the diagonal. The memory cost then boils down to the inversion for each harmonic $n$ of a linear system involving the shifted mean-Jacobian,  Hence, thanks to parallelization, the preconditioned GMRES algorithm allows to quickly perform resolvent analysis at a reasonable memory cost, since the limiting step corresponds to the inversion by each processor of a classical fixed-point Jacobian operator (a step that could also be made iterative and parallel for 3D configurations). 
Note, however, that a block-Jacobi preconditioner is only appropriate
for diagonally-dominant block-matrices, that is, when the harmonics of the base flow are weak. If this condition is not fulfilled, block Gauss-Seidel or more advanced, flexible inner-outer GMRES strategies combined with deflation techniques may also be considered \citep{jadoui2022comparative}.\\

\begin{centering}\subsubsection{Matrix-free implementation: adjoint-looping}\label{subsec:timestep}\end{centering}

Adjoint-looping has been proposed by Monokrousos \textit{et al.} \citep{monokrousos2010global} and later refined by Martini \textit{et al.} \citep{martini2021efficient} to compute the optimal forcing/response/gain in the LTI case. \textcolor{black}{By implementing randomized SVD directly in the time domain (RSVD-$\Delta t$) \cite{farghadan2025scalable}, Farghadan \textit{et al.} \cite{farghadan2024efficient} recently extended the framework to the harmonic resolvent setting}. A similar approach could be used to tackle the mean resolvent, by using harmonic input and harmonic averaging of the output, as explained in \ref{subsec:timedomain}.\\
\indent \, The matrix-based and matrix-free approaches have different strengths and weaknesses. The matrix-based approach is memory-intensive for large numbers of harmonics but easily parallelizable. The matrix-free paradigm has a low memory footprint but may be time-consuming, due to the sequential nature of adjoint-looping; parallelisation in time of direct-adjoint looping is possible but difficult \citep{skene2021parallel,costanzo2022parallel}.\\
\indent \, \textcolor{black}{In any case, as already mentioned in the Introduction, the goal of the present paper is not to benchmark the different implementations of adjoint-based methods for mean resolvent analysis. Rather, the focus is on developing a methodology that does not require the adjoint of the mean resolvent operator. This `adjoint-free' approach (only the adjoint of the mean-flow resolvent is needed) is presented in the next subsection. The adjoint-based approach is therefore used solely for validation, and only the linear algebra–based implementation is considered here.}\\

\textcolor{black}{\begin{centering}\subsection{Projection-based method for mean resolvent analysis}\label{eq:mainsec}\end{centering}}

Leclercq \& Sipp \citep{leclercq2023mean} showed that the resolvent operator about the mean flow approximates the mean resolvent. As recalled in \S~\ref{subsec:Sec1sub0}, this connection requires nearly quadratic nonlinearity (so $\mathsfbi{R}_{\overline{\mathbf{Q}}}\approx \mathsfbi{R}_{\overline{\mathsfbi{J}}}$) and weak unsteadiness (so $\mathsfbi{R}_{\overline{\mathsfbi{J}}}\approx\mathsfbi{R}_0$). This similarity may be leveraged to propose a projection method that eliminates the need to evaluate the adjoint mean resolvent operator.\\
\indent \, The method hinges on the assumption that the $d\ll \mathrm{N}$ leading mean-resolvent modes may be written as a superposition of the $d$ optimal mean-flow resolvent modes, i.e.
\begin{equation}
    \hat{\boldsymbol{\Psi}}_{MR}\approx\hat{\boldsymbol{\Psi}}_{MF}\boldsymbol{\Gamma}\label{eq:proj}
\end{equation}
where
\begin{equation}
    \hat{\boldsymbol{\Psi}}_{MF}=[\hat{\boldsymbol{\psi}}_{1,MF},\dots,\hat{\boldsymbol{\psi}}_{d,MF}],\quad
    \hat{\boldsymbol{\Psi}}_{MR}=[\hat{\boldsymbol{\psi}}_{1,MR},\dots,\hat{\boldsymbol{\psi}}_{d,MR}],
\end{equation}
are $\mathrm{N}\times d$ tall-and-skinny complex matrices and 
\begin{equation}
    \boldsymbol{\Gamma}=(\gamma_{ij})=[\boldsymbol{\gamma}_1,\dots,\boldsymbol{\gamma}_d]
\end{equation}
is a small $d\times d$ matrix of unknown projection coefficients. The generalised eigenvalue problem (\ref{eq:ghevp}) may be written in compact form for the $d$ leading mean resolvent modes as
\begin{equation}
\label{eq:allghevp}
    \mathsfbi{R}_0^H\mathsfbi{M}_y\mathsfbi{R}_0 \hat{\boldsymbol{\Psi}}_{MR} =\mathsfbi{M}_f\hat{\boldsymbol{\Psi}}_{MR}\boldsymbol{\Lambda}^2.
\end{equation}
Plugging (\ref{eq:proj}) into (\ref{eq:allghevp}) leads to
\begin{equation}
\label{eq:interm}
    \mathsfbi{R}_0^H\mathsfbi{M}_y\mathsfbi{R}_0\hat{\boldsymbol{\Psi}}_{MF} \boldsymbol{\Gamma}=\mathsfbi{M}_f\hat{\boldsymbol{\Psi}}_{MF} \boldsymbol{\Gamma} \boldsymbol{\Lambda}^2.
\end{equation}
The mean-flow resolvent modes are $\mathsfbi{M}_f$-orthogonal, i.e. $\hat{\boldsymbol{\Psi}}_{MF}^H\mathsfbi{M}_f$$\hat{\boldsymbol{\Psi}}_{MF}=\mathsfbi{I}$, therefore upon multiplying (\ref{eq:interm}) on the left by \textcolor{black}{$\hat{\boldsymbol{\Psi}}_{MF}^H$}, we obtain a small $d\times d$ eigenvalue problem 
\begin{equation}
    \hat{\boldsymbol{\Psi}}_{MF}^H\mathsfbi{R}_0^H\mathsfbi{M}_y\mathsfbi{R}_0\hat{\boldsymbol{\Psi}}_{MF}\boldsymbol{\Gamma}=\boldsymbol{\Gamma}\boldsymbol{\Lambda}^2,
\end{equation}
or equivalently,
\begin{equation}
\label{eq:smallghevp}
    \hat{\mathsfbi{Y}}^H\mathsfbi{M}_y\hat{\mathsfbi{Y}}\boldsymbol{\Gamma}=\boldsymbol{\Gamma}\boldsymbol{\Lambda}^2,\quad\text{with}\quad\hat{\mathsfbi{Y}}=\mathsfbi{R}_0\hat{\boldsymbol{\Psi}}_{MF}.
\end{equation}
The eigenvalues of this small problem are those of the initial problem, and the eigenvectors provide the unknown modal amplitudes. The columns of the skinny matrix $\hat{\mathsfbi{Y}}$ correspond to the Fourier components at $\omega$ of the linear responses to harmonic forcing by optimal mean-flow resolvent modes. These responses may be evaluated either using a matrix-based implementation (\S \ref{subsec:freqdomain}) or a matrix-free implementation (i.e. time-stepping, \S \ref{subsec:timedomain}). The computation of the $d$ responses may be carried out in parallel. Once the matrix is built, multiplication by $\hat{\mathsfbi{Y}}^H$ is a trivial task. Note that since we require that $\hat{\boldsymbol{\Psi}}_{MR}^H\mathsfbi{M}_f$$\hat{\boldsymbol{\Psi}}_{MR}=\mathsfbi{I}$, the matrix $\boldsymbol{\Gamma}$ must be unitary, i.e. $\boldsymbol{\Gamma}^H\boldsymbol{\Gamma}=\boldsymbol{\Gamma}\boldsymbol{\Gamma}^H=\mathsfbi{I}$; in other terms, eigenvectors of $\hat{\mathsfbi{Y}}^H\mathsfbi{M}_f\hat{\mathsfbi{Y}}$ forming the columns $\boldsymbol{\gamma}_j$ of $\boldsymbol{\Gamma}$ need to be normalized to 1. Once the small eigenvalue problem is solved, the optimal forcing modes are retrieved using (\ref{eq:proj}). The corresponding $\mathrm{N}\times d$ matrix of optimal responses solves 
\begin{equation}
\hat{\boldsymbol{\Phi}}_{MR}\boldsymbol{\Lambda}=\mathsfbi{R}_0 \hat{\boldsymbol{\Psi}}_{MR}
\end{equation}
so, using (\ref{eq:proj}) and (\ref{eq:smallghevp}), $\hat{\boldsymbol{\Phi}}_{MR}=\hat{\mathsfbi{Y}}\boldsymbol{\Gamma}\boldsymbol{\Lambda}^{-1}$ (we choose $d$ less than the rank $r$ of the mean resolvent so $\boldsymbol{\Lambda}$ is always invertible).\\
\indent \, The procedure is summarised in algorithm \ref{alg:Tab0}. 
To study the convergence of the approximation with respect to the subspace dimension $d$, we will later use a superscript $(.)^d$ when necessary.\\


\begin{myalgorithm}{Projection method for mean resolvent analysis of periodic flows.}
\label{alg:Tab0}

\begin{algorithmic}[1]

\STATE Construct the input subspace
$\hat{\boldsymbol{\Psi}}_{MF}
=
[\hat{\boldsymbol{\psi}}_{1,MF},\dots,\hat{\boldsymbol{\psi}}_{d,MF}]$
from $d$ optimal forcing mean-flow resolvent modes.

\STATE Compute (in parallel) the $d$ mean linear responses
$\hat{\mathsfbi{Y}}$ to these harmonic input modes:

\begin{itemize}
    \item either in the frequency domain, using the harmonic resolvent operator
    (see Sec.~\ref{subsec:freqdomain}),
    \item or in the time domain, by time-marching the LTP dynamics
    (\eqref{eq:eq3MR}-\eqref{eq:eq3bisMR}).
\end{itemize}

\STATE Solve reduced eigenvalue problem
\[
(\hat{\mathsfbi{Y}}^H\mathsfbi{M}_y\hat{\mathsfbi{Y}})
\boldsymbol{\gamma}_j
=
\lambda^2_{j,MR}\boldsymbol{\gamma}_j
\]
providing approximation of optimal energy gains
$\lambda^2_{j,MR}$.

\STATE Reconstruct mean resolvent optimal input and output modes as
\[
\hat{\boldsymbol{\psi}}_{j,MR}
=
\hat{\boldsymbol{\Psi}}_{MF}\boldsymbol{\gamma}_j,
\qquad
\hat{\boldsymbol{\phi}}_{j,MR}
=
\hat{\mathsfbi{Y}}\boldsymbol{\gamma}_j/\lambda_{j,MR}.
\]

\end{algorithmic}

\end{myalgorithm}


\begin{centering}\section{Mean-flow resolvent versus mean resolvent analysis: case of an axisymmetric time-periodic jet}\label{sec:Sec3}\end{centering}

In this section, we compare mean-flow resolvent and mean resolvent analyses using the adjoint-based matrix-based approaches described in \S~\ref{subsec:linalg}. The present section also provides reference results for the convergence analysis of the \textcolor{black}{projection approach} in \S~\ref{sec:Sec4sub1}. As a benchmark case, we select a problem similar to that investigated by Shaabani-Ardali, Sipp \& Lesshafft \citep{shaabani2019vortex} and Padovan \& Rowley \citep{padovan2022analysis}, i.e. an axisymmetric laminar jet forced by an axial inflow velocity oscillating periodically. As in these prior works, we will consider axisymmetric perturbations only.\\

\begin{centering}\subsection{Flow configuration and spatial discretisation}\label{subsec:Sec3sub1}\end{centering}

Although the flow studied in these prior works is strictly incompressible, the present numerical analysis is based on a compressible solver. The compressible Navier--Stokes equations formulated in the two-dimensional cylindrical coordinates $r$ and $z$, and expressed in conservative variables $\boldsymbol{\mathrm{q}}=\left\{\rho,\rho\boldsymbol{\mathrm{u}},\rho E\right\}^T$, read:
\begin{subequations}
\begin{equation}
\label{eq:eq18a}
\frac{\partial \rho}{\partial t}+\nabla\boldsymbol{\cdot}\left(\rho \boldsymbol{\mathrm{u}}\right)=0,
\end{equation}
\begin{equation}
\label{eq:eq18b}
\frac{\partial \left(\rho \boldsymbol{\mathrm{u}}\right)}{\partial t}+\nabla\boldsymbol{\cdot}\left(\rho\boldsymbol{\mathrm{u}}\otimes\boldsymbol{\mathrm{u}}\right)+\nabla p-\nabla\boldsymbol{\cdot}\boldsymbol{\tau}=0,
\end{equation}
\begin{equation}
\label{eq:eq18c}
\frac{\partial \left(\rho E\right)}{\partial t}+\nabla\boldsymbol{\cdot}\left[\left(\rho E+p\right)\boldsymbol{\mathrm{u}}-\boldsymbol{\tau}\boldsymbol{\cdot}\boldsymbol{\mathrm{u}}-\frac{1}{Re Pr}\frac{\nabla T}{\left(\gamma-1\right)M^2}\right]=0,
\end{equation}
\noindent \noindent with $\boldsymbol{\mathrm{u}}=\left\{u_r,u_z\right\}$, $\boldsymbol{\tau}=-\frac{2}{3Re}\left(\nabla\boldsymbol{\cdot}\boldsymbol{\mathrm{u}}\right)\mathsfbi{I}+\frac{1}{Re}\left(\nabla\boldsymbol{\mathrm{u}}+\nabla\boldsymbol{\mathrm{u}}^T\right)$ the viscous stress tensor for a Newtonian fluid, whereas $p$ and $T$ denote pressure and temperature, respectively. These equations are supplemented by the equations of state for a thermally and calorically perfect gas,
\begin{equation}
\label{eq:eq18d}
p=\frac{1}{\gamma M^2}\rho T,
\end{equation}
\begin{equation}
\label{eq:eq18e}
E=\frac{T}{\gamma\left(\gamma-1\right)M^2}+\frac{\|\boldsymbol{\mathrm{u}}\|^2}{2}.
\end{equation}
\end{subequations}
\noindent Eqs.~\eqref{eq:eq18a}-\eqref{eq:eq18e} have been made non-dimensional by using the inlet diameter $D$, the centreline velocity $U_c$, density $\rho_c$ and temperature $T_c$ at the inlet. The viscosity $\mu$ and the thermal conductivity $\kappa$ are assumed to be constant throughout the flow. The flow parameters defined in terms of dimensional quantities are the Reynolds, Mach and Prandtl numbers $Re=\rho_cU_cD/\mu$, $M=U_c/c_c$ (with $c_c$ the speed of sound on the centreline), $Pr=\mu c_p/\kappa$ (with $c_p$ the specific heat capacity at constant pressure), the ratio of ambient-to-jet temperature $S=T_{\infty} /T_c$, defined at the inlet, and the ratio of specific heat capacities $\gamma=c_p/c_v$. To compare with Padovan \& Rowley \cite{padovan2022analysis}, we select a Reynolds number $Re=1000$, and we impose a unitary temperature ratio $S=1$ with a Mach number $M=0.1$, for which compressibility effects are expected to be negligible. The adiabatic constant $\gamma$ is equal to 1.4, while the Prandtl number is assumed to be $Pr=1$ as in Ref.~\onlinecite{lesshafft2006nonlinear}.\\
\indent \, The governing Eqs.~\eqref{eq:eq18a}-\eqref{eq:eq18e} are discretised using the open-source CFD (Computational Fluid Dynamics) code BROADCAST presented in Ref.~\onlinecite{poulain2023broadcast} and which deals with the compressible Navier--Stokes equations within a finite-volume framework. The viscous fluxes are computed on a five-point compact stencil (fourth-order accurate) \citep{shen2009high}, while the space discretisation for the inviscid flux follows the FE-MUSCL scheme (Flux-Extrapolated-MUSCL) \citep{cinnella2016high,sciacovelli2021assessment}, available in different orders of accuracy; here, a fifth-order scheme was used. Linearisation of the governing Eqs.~\eqref{eq:eq18a}-\eqref{eq:eq18e} about a given reference state is performed with algorithmic differentiation using the open library TAPENADE \citep{hascoet2013tapenade}.\\
\indent \, Symmetry conditions $\partial\left(\rho, \rho u_z, \rho E\right)/\partial r = 0$, $\rho u_r=0$ are imposed at $r = 0$ by mirroring the values of the flow variables onto three ghost points across the axis. At the outlet and top boundaries, respectively, a zero\textit{th}-order extrapolation condition and a no-reflection condition with prescribed far-field are enforced. Lastly, and similarly to Ref.~\onlinecite{lesshafft2006nonlinear}, at the inlet, we apply a non-reflecting characteristic boundary condition with incoming characteristics:
\begin{subequations}
\begin{equation}
\label{eq:eq19a}
u_z\left(r\right)=g\left(t\right)\left[\frac{1}{2}-\frac{1}{2}\tanh{\frac{1}{4\theta_0}\left(r-\frac{1}{4r}\right)}\right],\ \ \ \ \ u_r\left(r\right)=0,
\end{equation}
\begin{equation}
\label{eq:eq19b}
T\left(r\right)=S+\left(1-S\right)u_z\left(r\right)+\frac{\gamma-1}{2}M^2u_z\left(r\right)\left(1-u_z\left(r\right)\right),
\end{equation}
\begin{equation}
\label{eq:eq19c}
\rho\left(r\right)=T^{-1}\left(r\right),
\end{equation}
\end{subequations}
\noindent where the axial inlet velocity $u_z\left(r\right)$ is modulated periodically by $g\left(t\right)=1+A\cos{\omega_0 t}$. The nondimensional vorticity thickness of the incoming profile and the amplitude of the periodic modulation are set, respectively, to $\theta_0=0.025$ and $A=0.05$ as in Ref.~\onlinecite{padovan2020analysis}. We will consider two different angular frequencies for the base flow, specifically, $\omega_0=6\pi/5$ as in Ref.~\onlinecite{padovan2022analysis} and half this value, $\omega_0=3\pi/5$. The two cases will be referred to as the `weakly unsteady' and `strongly unsteady' cases, respectively, as will become clear in the next subsection. Because compressibility effects are weak ($M=0.1$), the numerical tools are quantitatively validated for the case $\omega_0=6\pi/5$ against the incompressible results of Ref.~\onlinecite{padovan2022analysis} (see App.~\ref{sec:AppB}).\\
\indent \, While the full numerical domain is defined as the region $0\leq z\leq 30$ and $0\leq r\leq 10$, the `physical domain' is restricted to $D=\left\{\left(r,z\right)|~0\le r\le 2.5, 0\le z \le 20\right\}$, consistent with Ref.~\onlinecite{padovan2022analysis}. Grid-stretching is used to create sponge-like regions for $z\geq 20$ and $r\geq 2.5$ as in Ref.~\onlinecite{lesshafft2006nonlinear}, in order to mitigate reflections at the boundaries. Within the physical domain, the output matrix $\mathsfbi{C}$ is defined as a restriction operator, extracting the two momentum components of the state vector for mesh points falling within $D$. Conversely, the input matrix $\mathsfbi{B}=\mathsfbi{C}^T$ extends a momentum field defined over the physical domain $D$ to a full state vector defined over the entire numerical domain, by prolongating the input vector with zero entries. Within the physical domain, the mesh is uniform in the $r$ and $z$ directions respectively, with a grid size of $0.95\mathrm{N_r}\times 0.95\mathrm{N_z}=190\times 285$ (totaling $\mathrm{N}=4\times \mathrm{N_r} \times \mathrm{N_z}=240,000$ degrees of freedom) providing a satisfactory precision/cost trade-off, as indicated in App.~\ref{sec:AppB}.\\
\indent \, For resolvent analyses, the input norm will measure the $L_2$ norm of the momentum forcing within $D$, while the output norm will correspond to the $L_2$ norm of the momentum response within $D$. Therefore, we define the inner product matrix $\mathsfbi{M}_y=\mathsfbi{M}_f=\mathsfbi{M}$ as a diagonal matrix containing the surface of the cells in $D$, twice (for the two components of the input/output vector). Because the density field is almost constant and equal to $\rho\approx 1$ (weak compressibility $M=0.1$ and temperature ratio $S=1$), the squared output norm approximates the kinetic energy of the linear response within $D$, up to a factor of 2. The inner product matrix $\mathsfbi{M}$ will also be used to compute the energy in $D$ of the Fourier components of the periodic base flow.\\

\begin{centering}\subsection{Periodic base flow and mean flow}\label{subsec:Sec3sub2}\end{centering}

Shaabani-Ardali, Sipp \& Lesshafft \citep{shaabani2019vortex} studied the stability of the problem and showed that for $Re\lesssim 1350$ the system admits a stable periodic solution. We compute this base flow $\boldsymbol{\mathrm{Q}}\left(t\right)$ by time-marching~\eqref{eq:wqBF} with a standard 4th-order Runge--Kutta method. We then sample the permanent regime,
$\boldsymbol{\mathrm{Q}}\left(t_k;\phi\right)$, at $2\mathrm{N}_{\mathrm{h}}+1=33$ discrete Fourier collocation points in time defined by $t_k=k\Delta t$ with $\Delta t=T_0/ \left(2\mathrm{N}_{\mathrm{h}} + 1\right)$ and $0\leq k \leq 2\mathrm{N}_\mathrm{h}$. An instantaneous Jacobian $\mathsfbi{J}_{\boldsymbol{\mathrm{Q}}}\left(t_k;\phi\right)$ associated with each of these states is also evaluated using algorithmic differentiation \citep{poulain2023broadcast}. These snapshots are then used to decompose the periodic base flow and Jacobian into their (truncated) Fourier series coefficients by applying a Discrete Fourier Transform (DFT):
\begin{equation}
\label{eq:eq21}
\hat{\boldsymbol{\mathrm{Q}}}_n=\frac{1}{2\mathrm{N}_{\mathrm{h}}+1}\sum_{k=0}^{2\mathrm{N}_{\mathrm{h}}}\boldsymbol{\mathrm{Q}}\left(t_k;\phi\right)\mathrm{e}^{-\text{i}\frac{2\pi k n}{2\mathrm{N}_{\mathrm{h}}+1}},\ \ \ \ \ \ \hat{\mathsfbi{J}}_n=\frac{1}{2\mathrm{N}_{\mathrm{h}}+1}\sum_{k=0}^{2\mathrm{N}_{\mathrm{h}}}\mathsfbi{J}_{\boldsymbol{\mathrm{Q}}}\left(t_k;\phi\right)\mathrm{e}^{-\text{i}\frac{2\pi k n}{2\mathrm{N}_{\mathrm{h}}+1}},
\end{equation}
\noindent with $-\mathrm{N}_{\mathrm{h}}\leq n\leq \mathrm{N}_{\mathrm{h}}$. The mean field $\overline{\boldsymbol{\mathrm{Q}}}=\hat{\boldsymbol{\mathrm{Q}}}_0$ is used to construct $\mathsfbi{J}_{\overline{\boldsymbol{\mathrm{Q}}}}$ for mean-flow resolvent analysis, whereas the various $\hat{\mathsfbi{J}}_n$ are needed to evaluate the Hill matrix \eqref{eq:Hill} involved in the mean resolvent operator. Note that since the governing equations are nearly quadratic (weak compressibility effects), we are in a situation where $\mathsfbi{R}_{\overline{\mathsfbi{J}}}\approx \mathsfbi{R}_{\overline{\mathbf{Q}}}$ so deviations between $\mathsfbi{R}_{\overline{\mathbf{Q}}}$ and $\mathsfbi{R}_0$ are solely due to the unsteadiness of the base flow.

\begin{figure}[t!]
\centering
\includegraphics[width=1\textwidth]{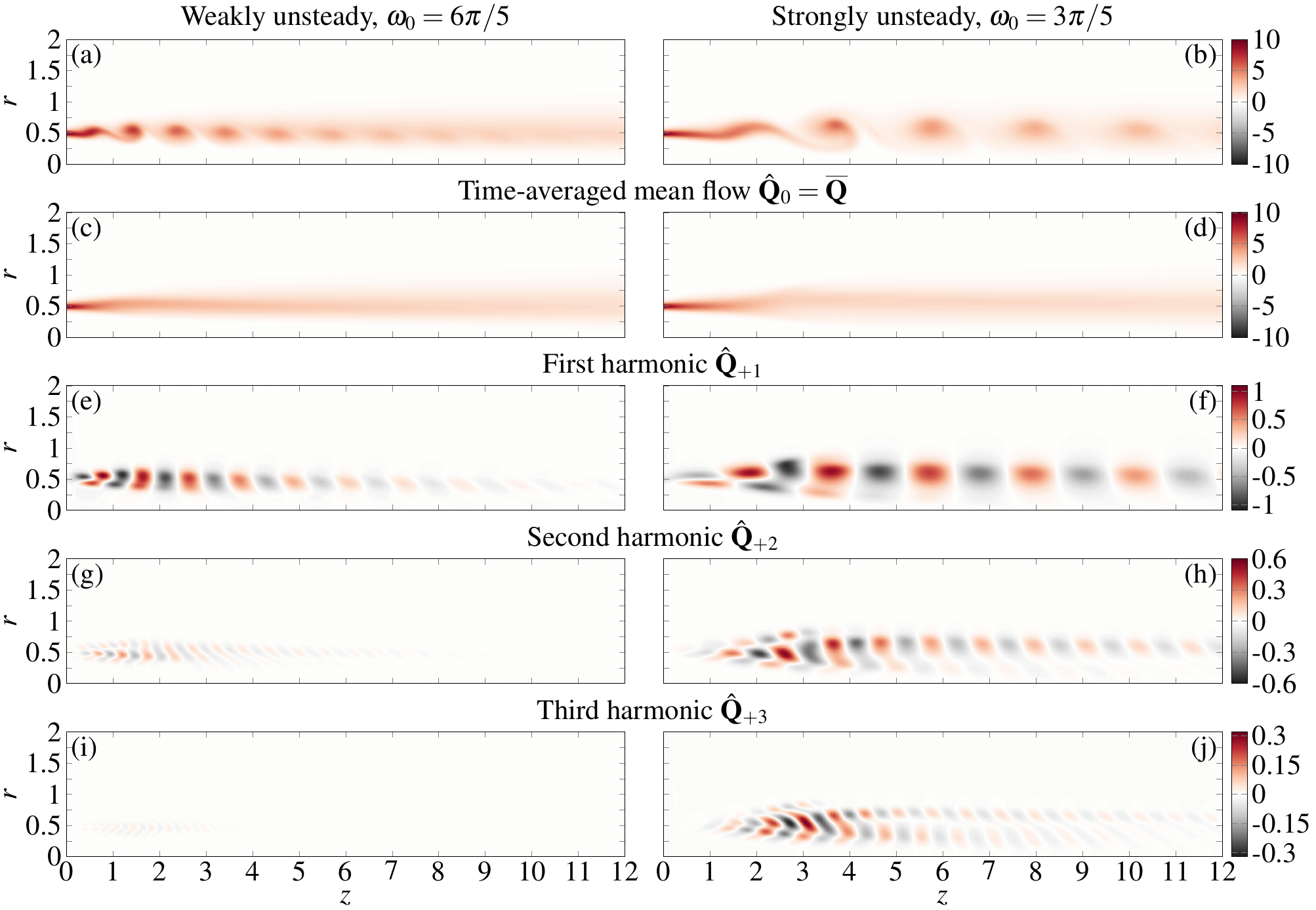}
\caption{\textcolor{black}{(a)-(b) Snapshots of the azimuthal vorticity field computed from the $T_0$-periodic base flow solution $\boldsymbol{\mathrm{Q}}\left(t\right)$ for (a) $\omega_0=6\pi/5$ and (b) $\omega_0=3\pi/5$. (c)-(d) Time-averaged azimuthal vorticity field from the corresponding mean state $\overline{\boldsymbol{\mathrm{Q}}}$. (e)-(f) First $n=1$, (g)-(h) second $n=2$ and (i)-(j) third $n=3$ harmonics extracted from the Fourier decomposition of the instantaneous flow, shown as the real part of the azimuthal vorticity field. The same mesh and Reynolds number $Re=1000$ were used in both computations.}} 
\label{fig:Fig12} 
\end{figure}

\begin{figure}[t!]
\centering
\includegraphics[width=1\textwidth]{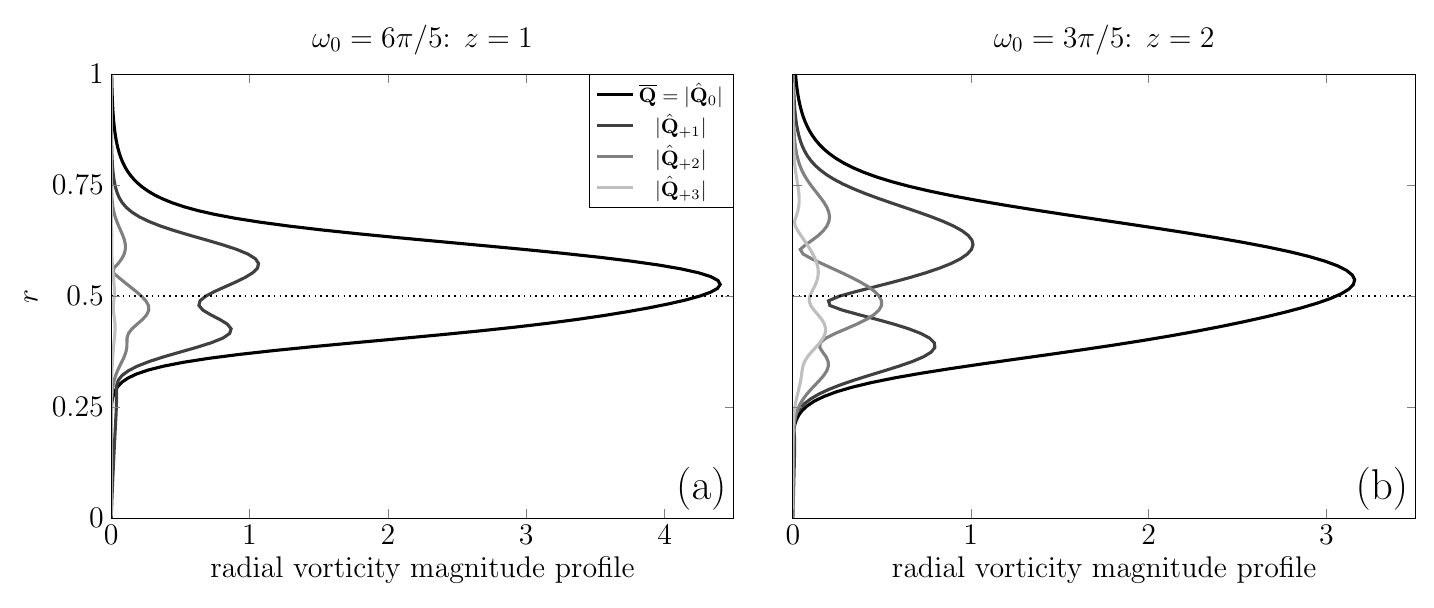}
\caption{Radial vorticity magnitude profile associated with the mean flow $\overline{\boldsymbol{\mathrm{Q}}}=\hat{\boldsymbol{\mathrm{Q}}}_0$ and the first three harmonics $\hat{\boldsymbol{\mathrm{Q}}}_{n}$ ($n=1,2,3$) extracted from the Fourier decomposition of the instantaneous flow (see Fig.~\ref{fig:Fig12}) for (a) $\omega_0=6\pi/5$ at an axial coordinate $z=1$ and (b) $\omega_0=3\pi/5$ at $z=2$.}
\label{fig:FigN1} 
\end{figure}

\begin{figure}[t!]
\centering
\includegraphics[width=0.8\textwidth]{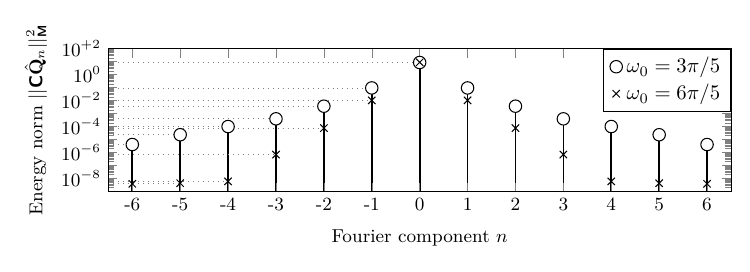}
\caption{Energy norm of the various harmonics ($|n|\leq 6$) extracted from the Fourier decomposition of the two $T_0$-periodic base flow solution $\boldsymbol{\mathrm{Q}}\left(t\right)$ for $\omega_0=3\pi/5$ (empty circle) and $6\pi/5$ (black crosses).}
\label{fig:FigN2} 
\end{figure}

Two snapshots of the azimuthal vorticity magnitude from the $T_0$-periodic base flow solution $\boldsymbol{\mathrm{Q}}\left(t\right)$ are shown in Fig.~\ref{fig:Fig12}(a) and (b), respectively for $\omega_0=6\pi/5$ and $3\pi/5$, with their associated mean flow fields $\overline{\boldsymbol{\mathrm{Q}}}$ illustrated in panels~\ref{fig:Fig12}(c) and (d). The real parts of the first three harmonics extracted from the two base flows are shown in panels (e-j), together with the radial profiles extracted at a prescribed $z$ coordinate displayed in Fig.~\ref{fig:FigN1}. In Fig.~\ref{fig:FigN2}, we report $\|\mathsfbi{C}\hat{\mathbf{Q}}_n\|_{\mathsfbi{M}}^2$ for each of these Fourier components, which approximates their kinetic energy in $D$. It appears from Fig.~\ref{fig:Fig12}(b) that the base flow configuration oscillating at $\omega_0=3\pi/5$ substantially differs from the one of Fig.~\ref{fig:Fig12}(a) for the size of the coherent structures and wavelength between two consecutive vortices, which, consistently with the imposed forcing frequency, appears as approximately twice that of the first jet (vortex-paired configuration). In the following, we will designate the $\omega_0=6\pi/5$ and $3\pi/5$ base flow configurations as `weakly unsteady' and `strongly unsteady', respectively. This distinction is made based on the frequency content of Fig.~\ref{fig:FigN2}. In the weakly unsteady case, the dynamics is well approximated by the mean and first harmonic, with the second harmonic only weakly participating and higher harmonics being completely negligible. On the contrary, in the strongly unsteady case, a higher number of harmonics must be accounted for to reconstruct the flow. It is also interesting to comment on the radial structure of these higher harmonics: as visible in Fig.~\ref{fig:FigN1}, they display a radial modulation, showing two or more lobes, a feature that will leave a footprint on mean resolvent modes as we shall see later.\\

\begin{centering}\subsection{Mean-flow resolvent analysis}\label{subsec:Sec3sub3}\end{centering}

\begin{figure}[t!]
\centering
\includegraphics[width=0.9\textwidth]{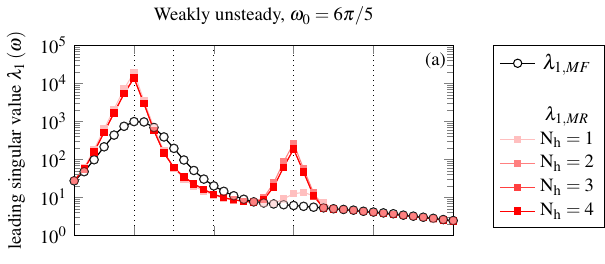}\\
\includegraphics[width=0.9\textwidth]{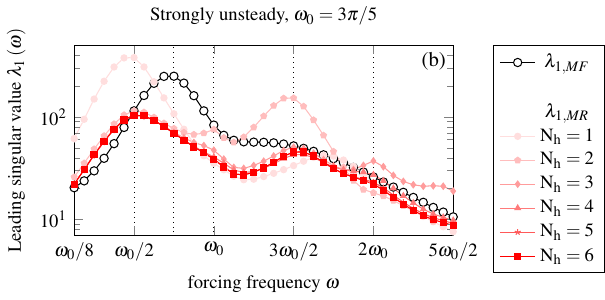}\\
\caption{(a) Leading singular value $\lambda_1$, as a function of the forcing frequency $\omega$, for the weakly unsteady base flow oscillating at $\omega_0=6\pi/5$. Black line with white circles: resolvent analysis about the mean flow. Red filled squares: mean resolvent analysis computed for $\mathrm{N}_{\mathrm{h}}=1,2,3$ and $4$. A good convergence is achieved already for $\mathrm{N}_{\mathrm{h}}=2$. A mesh of size $\mathrm{N_r}\times \mathrm{N_z}=200\times 300$ was used for this calculation. (b) Same, but for the strongly unsteady base flow oscillating at $\omega_0=3\pi/5$, for which convergence is achieved starting from $\mathrm{N}_h\ge 5$, at least in the explored frequency range.}
\label{fig:Fig5} 
\end{figure}

The mean-flow resolvent gain curve is reported in Fig.~\ref{fig:Fig5}(a,b) using white circles. We note that the results obtained for $\omega_0=6\pi/5$ (panel (a)) show a clear sub-harmonic amplification in agreement with Ref.~\onlinecite{padovan2022analysis} (see also App.~\ref{sec:AppB}); this is explained by the sensitivity of the flow to undergo vortex pairing. The mean flow analysis for the strongly unsteady base flow predicts instead a dominant peak occurring at $\omega \approx 3\omega_0/4$ (with $\omega_0 = 3\pi/5$), which seems physically irrelevant.\\

\begin{centering}\subsection{Mean resolvent analysis}\label{subsec:Sec3sub5}\end{centering}

The weakly and strongly unsteady cases are discussed in separate subsections below.\\

\begin{centering}\subsubsection{Weakly unsteady base flow}\label{subsubsec:SSS1}\end{centering}

\begin{figure}[t!]
\centering
\includegraphics[width=1\textwidth]{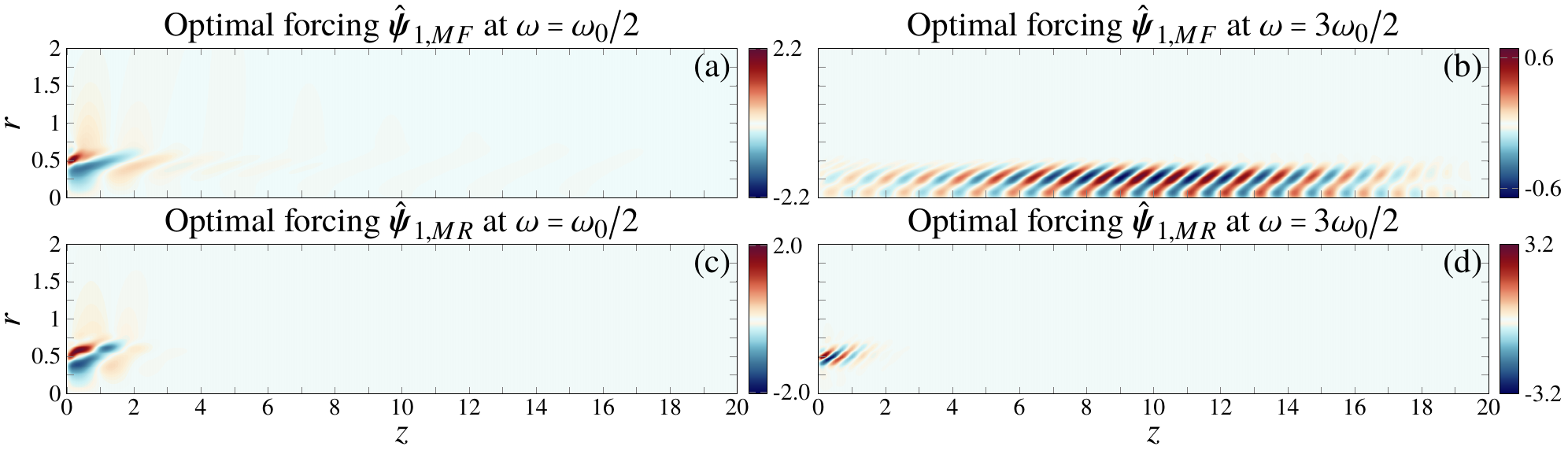}
\caption{Real part of the axial velocity component of the optimal forcing mode, $\hat{\boldsymbol{\psi}}_1$, computed according to (a,b) the resolvent analysis about the mean flow and (c,d) the mean resolvent analysis of the weakly unsteady case with $\omega_0=6\pi/5$, at (a,c) $\omega=\omega_0/2$ and (b,d) $\omega=3\omega_0/2$ (for $\mathrm{N}_h=2$). To `synchronise' mean-flow resolvent and mean resolvent modes, their phase is fixed to zero at $\left(r,z\right)=\left(0.5,0.1\right)$.}
\label{fig:Fig6} 
\centering
\bigskip
\includegraphics[width=1\textwidth]{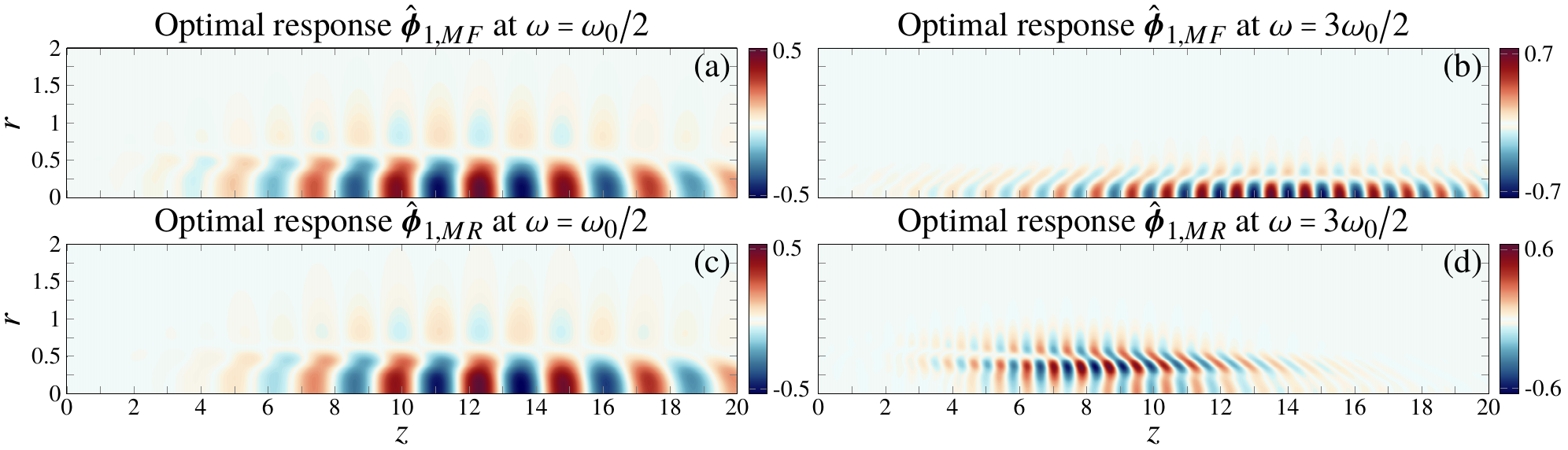}
\caption{Same caption as Fig.~\ref{fig:Fig6}, for the corresponding optimal response modes.}
\label{fig:Fig7} 
\end{figure}

The leading mean resolvent gain (red markers) is compared to the mean-flow gain curve in Fig.~\ref{fig:Fig5}(a). A convergence test with respect to the number of Fourier components indicates that $\mathrm{N}_{\mathrm{h}}=2$ is already sufficient for the weakly unsteady case. Similarly to the harmonic resolvent analysis discussed in Ref.~\onlinecite{padovan2022analysis} (see also App.~\ref{sec:AppB}), the mean resolvent predicts a much stronger amplification of sub-harmonic disturbances at $\omega\approx\omega_0/2$, which stems from interactions of the perturbation with the unsteady part of the base flow overlooked by the mean flow analysis. A well-defined secondary receptivity peak is also visible at $\omega\approx 3\omega_0/2$. For larger frequencies, the gain curve collapses on that predicted by the mean flow analysis, confirming that the base flow unsteadiness is relatively weak, i.e. only the interactions of the external forcing with the temporal mean, and those between the external forcing and the first two harmonics of the periodic base flow are significant.\\
\indent The optimal forcing modes computed at the two peak frequencies $\omega=\omega_0/2$ and $3\omega_0/2$ according to the mean-flow resolvent and mean resolvent analyses are displayed in Fig.~\ref{fig:Fig6}. Both mean-flow resolvent and mean resolvent modes capture structures aligned against the shear, indicative of non-modal amplification through the Orr mechanism \citep{orr1907stability}. At $\omega=\omega_0/2$, the optimal forcing mode of the mean resolvent has a radial modulation which is absent from its mean-flow counterpart. The pattern has two lobes in the radial direction, whereas the mean-flow mode only has one. This radial pattern is reminiscent of that of $\hat{\mathbf{Q}}_{\pm1}$ in Fig.~\ref{fig:FigN1}(a), which plays a role only in mean resolvent analysis. Discrepancies are even more pronounced at $3\omega_0/2$, where the mean resolvent mode is localised upstream in a tiny zone near the inlet, whereas the mean-flow mode is widely spread along the axial direction. The corresponding optimal responses are shown in Fig.~\ref{fig:Fig7}. At $\omega_0/2$, the mean-flow resolvent and mean resolvent modes are quite similar, whereas at $3\omega_0/2$ the mean resolvent mode support is more localised and centred more upstream than its mean-flow counterpart. 

\begin{figure}[t!]
\centering
\includegraphics[width=1\textwidth]{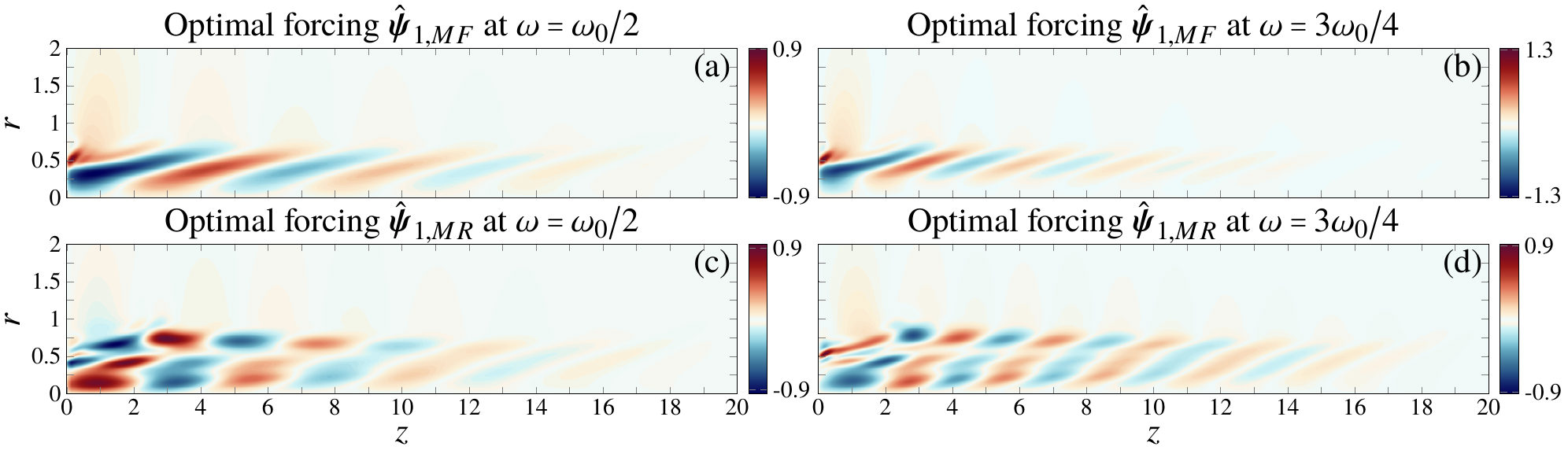}
\caption{Real part of the axial velocity component of the optimal forcing mode, $\hat{\boldsymbol{\psi}}_1$, computed according to (a,b) the resolvent analysis about the mean flow and (c,d) the mean resolvent analysis of the strongly unsteady case with $\omega_0=3\pi/5$, at (a,c) $\omega=\omega_0/2$ and (b,d) $\omega=3\omega_0/4$ (for $\mathrm{N}_h=3$). To `synchronize' mean-flow resolvent and mean resolvent modes, their phase is fixed to zero at $\left(r,z\right)=\left(0.5,0.1\right)$.} 
\label{fig:Fig17}
\centering
\bigskip
\includegraphics[width=1\textwidth]{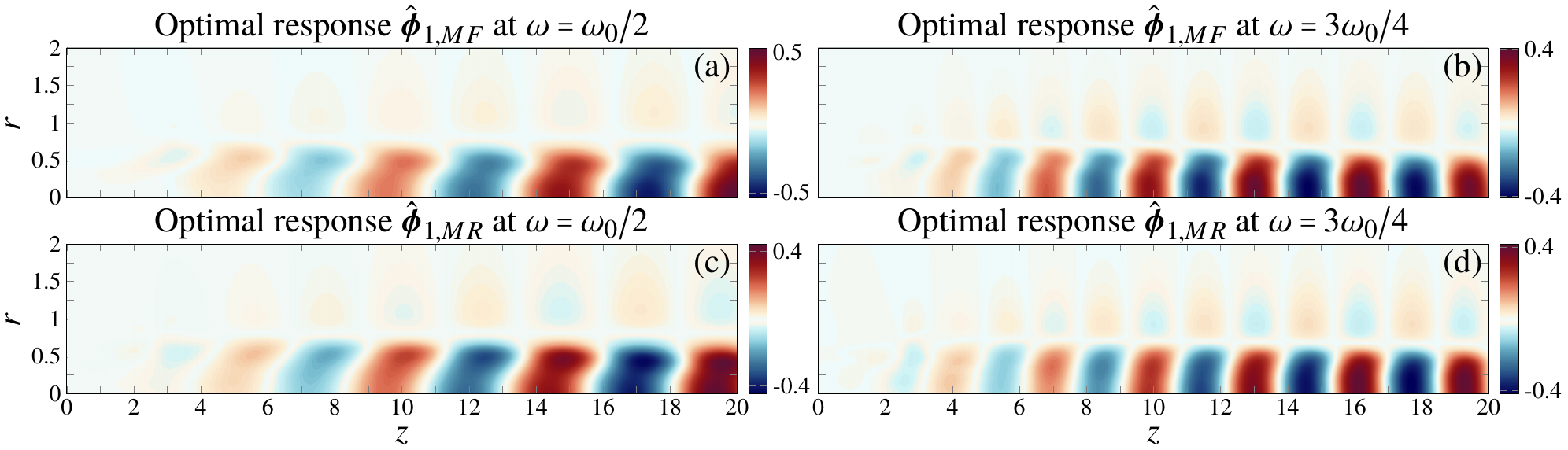}
\caption{\textcolor{black}{Same caption as Fig.~\ref{fig:Fig17}, for the corresponding optimal response modes.}} 
\label{fig:Fig17bis} 
\end{figure}

\indent To quantify the alignment between two modes $\mathbf{a}$ and $\mathbf{b}$, we compute the usual coefficient $0\leq \gamma_{\langle \mathbf{a},\mathbf{b}\rangle}\leq 1$, 
\begin{equation}
\label{eq:eq22bis}
\gamma_{\langle \mathbf{a},\mathbf{b}\rangle}=\frac{|\langle\boldsymbol{\mathrm{a}},\boldsymbol{\mathrm{b}}\rangle|}{\|\boldsymbol{\mathrm{a}}\| \|\boldsymbol{\mathrm{b}}\|}
\end{equation}
based on the inner product matrix $\mathsfbi{M}$. Numerical values are reported in table~\ref{tab:Tab1} for both input and output modes. At $\omega=\omega_0/2$, the alignment is greater than 0.85 for both the input and output modes, but at $\omega=3\omega_0/2$, the alignment is extremely poor at less than 0.1 for both input and output modes.\\

\begin{centering}\subsubsection{Strongly unsteady base flow}\label{subsubsec:SSS2}\end{centering}

Figure~\ref{fig:Fig5}(b) shows slower convergence of the leading mean resolvent optimal gain, compared to the weakly unsteady case: a truncation at $\mathrm{N}_{\mathrm{h}}\gtrsim 5$ is now necessary to ensure convergence in the whole frequency range, which confirms the strongly unsteady character of the $\omega_0=3\pi/5$ case. As a result, deviations between the mean-flow resolvent and mean resolvent analyses are stronger than for the weakly unsteady case. The mean resolvent operator correctly identifies the receptivity peak at $\omega_0/2$ associated with vortex pairing, but the mean-flow resolvent peaks at an intermediate frequency $\omega\approx 3\omega_0/4$, which seems physically irrelevant. As in the weakly unsteady case, the mean resolvent gain curve has a peak at $3\omega_0/2$, which is absent from the mean-flow counterpart. But contrary to the weakly unsteady case, the maximum amplification is now less for the mean resolvent than for the mean-flow resolvent. This denotes weaker sensitivity to subharmonic amplification, i.e. to induce vortex pairing, one must now provide greater input energy.

The optimal input and output structures computed according to the two resolvent analyses are shown in Figs.~\ref{fig:Fig17} and~\ref{fig:Fig17bis} respectively, at $\omega_0/2$ and $3\omega_0/4$. As in the weakly unsteady case, both operators capture the Orr mechanism, but the mean-flow resolvent fails at capturing the radial pattern of the optimal forcing mode. For the strongly unsteady case, there are now three distinct lobes in the radial direction, reminiscent of the structure of $\hat{\mathbf{Q}}_{\pm2}$ in Fig.~\ref{fig:FigN1}(b); again, a sign that base flow unsteadiness leaves a noticeable footprint on the optimal forcing mode. The optimal forcing mode of the mean resolvent has stronger receptivity close to the axis compared to the corresponding mean-flow mode. The spatial support of the optimal forcing mode of the mean resolvent is also more extended downstream at $\omega=3\omega_0/4$. All these differences lead to a poor alignment coefficient of less than 0.5 for the optimal forcing modes at the two frequencies. Despite this, the optimal response modes look very similar in~\ref{fig:Fig17bis} and have an alignment coefficient greater than 0.95 at the two frequencies.

\begin{table}[t!]
\centering
\begin{tabular}{cccccccc|cccccccc}
\multicolumn{8}{c}{weakly unsteady $\omega_0=6\pi/5$ } & \multicolumn{8}{c}{strongly unsteady $\omega_0=3\pi/5$ }  \\ \hline
$\omega$ & & $\gamma_{\langle\hat{\boldsymbol{\psi}}_{1,MF},\, \hat{\boldsymbol{\psi}}_{1,MR}\rangle}$  & & & $\gamma_{\langle\hat{\boldsymbol{\phi}}_{1,MF},\, \hat{\boldsymbol{\phi}}_{1,MR}\rangle}$ & & & & $\omega$ & & $\gamma_{\langle\hat{\boldsymbol{\psi}}_{1,MF},\, \hat{\boldsymbol{\psi}}_{1,MR}\rangle}$  & & & $\gamma_{\langle\hat{\boldsymbol{\phi}}_{1,MF},\, \hat{\boldsymbol{\phi}}_{1,MR}\rangle}$ & \\ \hline
$\omega_0/2$ & & 0.8501 & & & 0.9934 & & & & $\omega_0/2$ & & 0.4971 & & & 0.9510 & \\
$3\omega_0/2$ & & 0.0103 & & & 0.0732 & & & & $3\omega_0/4$ & & 0.4699 & & & 0.9911 & \\
\end{tabular}
\caption{Alignment coefficients \eqref{eq:eq22bis} between optimal input-output structures predicted by the two resolvent analyses, in both the weakly and strongly unsteady cases.}
\label{tab:Tab1}
\end{table}

As a final note, we underline the fact that the level of uncertainty associated with any LTI model increases with the level of base flow unsteadiness. But this limitation applies to both the mean-flow resolvent and the mean resolvent operators, because they both belong to the LTI framework.\\


\textcolor{black}{\begin{centering}\section{Mean-resolvent analysis: convergence analysis of the projection-based approach}\label{sec:Sec4sub1}\end{centering}}

\textcolor{black}{In the present section, we analyse the convergence of the projection-based approach against the results from \S \ref{sec:Sec3}}. Here we conveniently use the matrix-based implementation to evaluate the response at $\omega$ to harmonic forcing by mean-flow resolvent input modes, since we already worked with the harmonic resolvent in the previous section. But since the true benefit of the projection method arises when it is not possible or impractical to use a matrix-based approach, we also verify in \S~\ref{subsubsec:AppAsub1} that the time-stepping approach leads to identical results.\\

\begin{figure}[t!]
\centering
\includegraphics[width=0.9\textwidth]{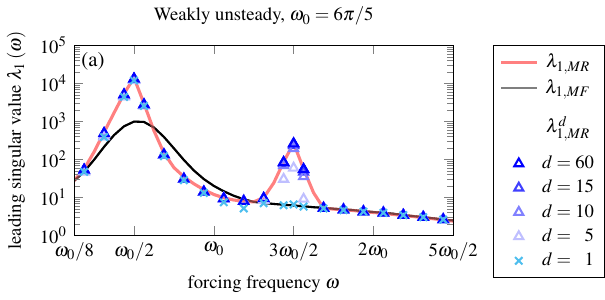}
\bigskip
\includegraphics[width=0.8\textwidth]{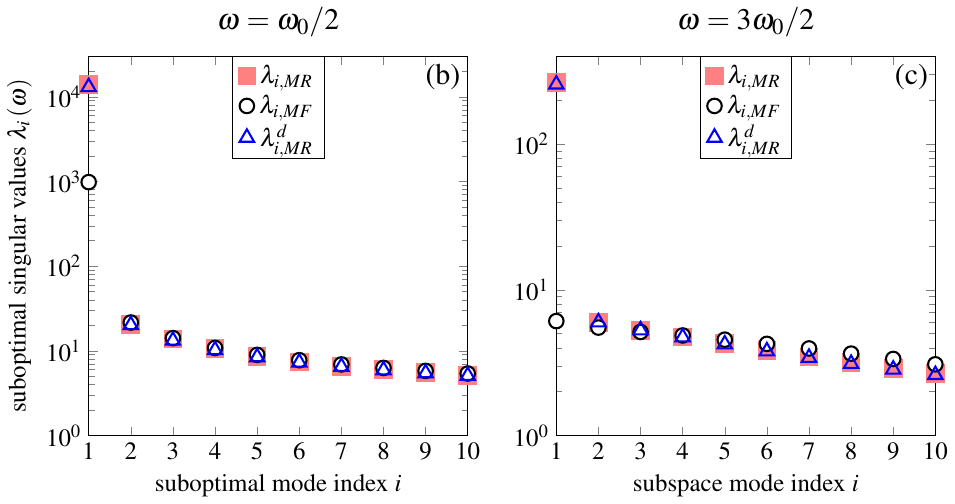}
\caption{(a) Leading singular value $\lambda_1$, as a function of the forcing frequency $\omega$, for the weakly unsteady case with $\omega_0=6\pi/5$. Black line: resolvent analysis about the mean flow. Red line: mean resolvent analysis computed for $\mathrm{N}_{\mathrm{h}}=2$. Blue empty triangles: optimal gain predicted by the \textcolor{black}{projection method} discussed in \S\ref{eq:mainsec} using increasing subspace dimensions up to $d=60$. The special case of a rank-one projection is depicted as light-blue crosses. (b) First 10 suboptimal singular values $\lambda_{i}\left(\omega\right)$ associated with the three analyses of panel (a) for $\omega=\omega_0/2$ and $d=60$. (c) Same as in (b) but for $\omega=3\omega_0/2$.}
\label{fig:Fig8} 
\end{figure} 

\begin{centering}\subsection{Weakly unsteady base flow}\label{subsec:SSSS1}\end{centering}

\begin{figure}[t!]
\centering
\begin{minipage}{0.45\textwidth}
\centering
\includegraphics[width=0.9\textwidth]{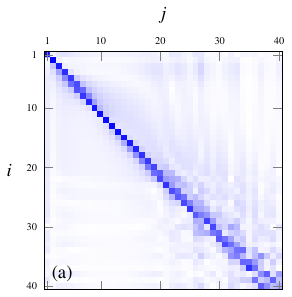}
\end{minipage}
\begin{minipage}{0.45\textwidth}
\centering
\includegraphics[width=0.985\textwidth]{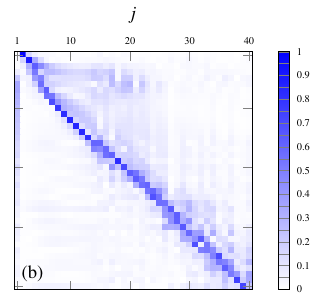}
\end{minipage}
\caption{Matrix of projection coefficients $\boldsymbol{\Gamma}=(\gamma_{ij})$ computed by solving~\eqref{eq:smallghevp} for (a) $\omega=\omega_0/2$ and (b) $\omega=3\omega_0/2$ (with $\omega_0=6\pi/5$) for $d=60$; for clarity, only the first $40\times40$ elements are shown.}
\label{fig:Fig9} 
\centering
\includegraphics[width=1\textwidth]{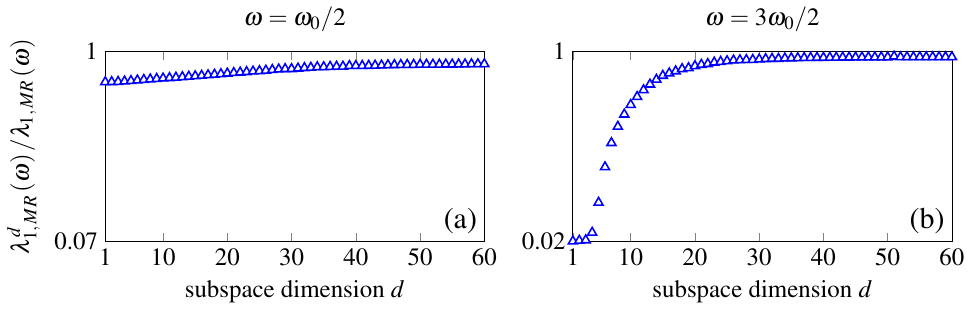}
\caption{Convergence of the leading mean resolvent singular value computed \textit{via} the projection method for (a) $\omega=\omega_0/2$ and (b) $\omega=3\omega_0/2$ (with $\omega_0=6\pi/5$). The values are rescaled by the converged value from the reference adjoint-based framework of \S\ref{sec:Sec3}, which is $13925.4$ in (a) and $264.7$ in (b). The lowest vertical axis limit has been selected to indicate the optimal gain from the corresponding mean flow analysis, which is $982.3$ in (a) and $6.1$ in (b). For $\omega=\omega_0/2$, the value of $\lambda^{d}_{1,\,MR}$ obtained with a single-mode ($d=1$) is $11845.9$ (approximately $85$\% of $\lambda_{1,\,MR}$), while a value of $13071.7$ is obtained for $d=60$ ($94$\% of $\lambda_{1,\,MR}$). Analogously, for $\omega=3\omega_0/2$, we have $6.5$ for $d=1$ (approximately $2.4$\% of $\lambda_{1,\,MR}$) and $257.18$ for $d=60$ ($97$\% of $\lambda_{1,\,MR}$).}
\label{fig:Fig10} 
\end{figure}

\begin{figure}[t!]
\centering
\includegraphics[width=1\textwidth]{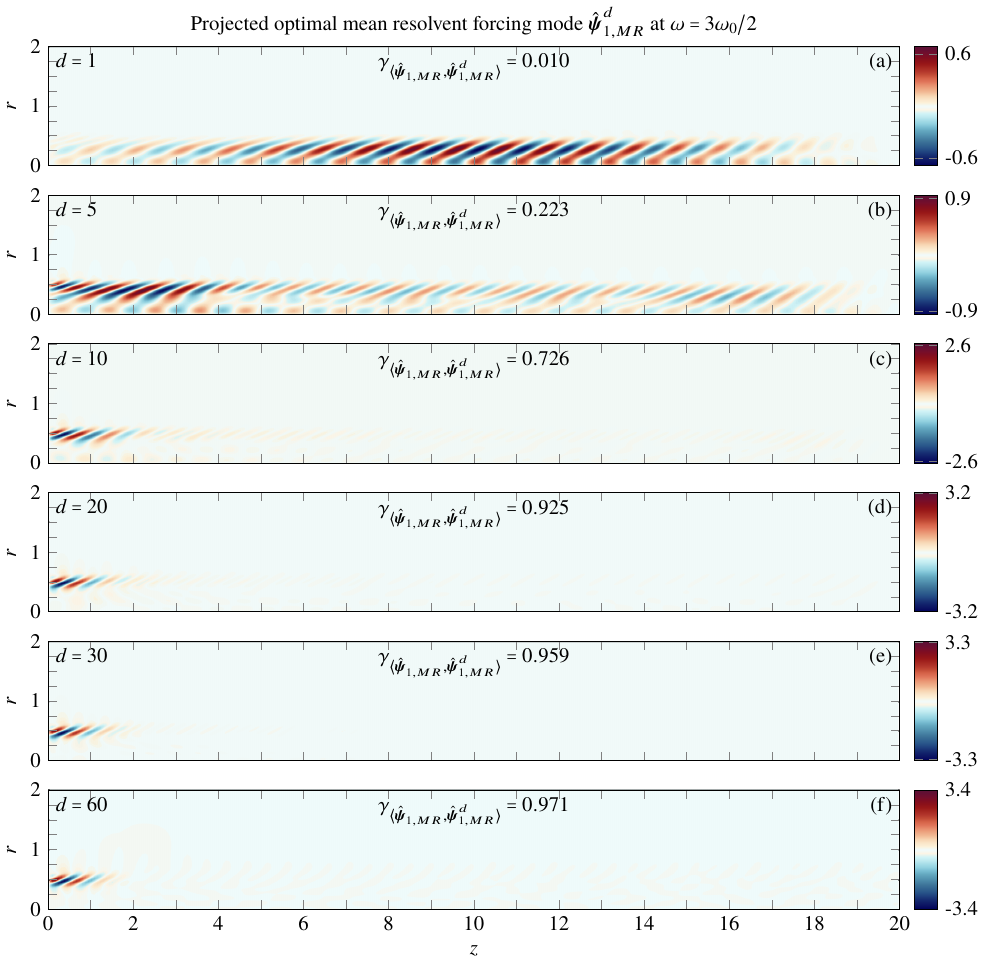}
\caption{Real part of the axial velocity component of the mean resolvent optimal forcing mode $\hat{\boldsymbol{\psi}}^d_{1,MR}$, approximated with the projection method, in the weakly unsteady case, using increasing subspace dimensions $d$: (a) 1, (b) 5, (c) 10, (d) 20, (e) 30 and (f) 60. To ease the visual comparison, each mode has been normalised such that its phase at a coordinate $\left(r,z\right)=\left(0.5,0.1\right)$ is zero. We also give the value of the alignment coefficients between the reference case $\hat{\boldsymbol{\psi}}_{1,MR}$ (not shown here) and the approximated one. }
\label{fig:Fig11} 
\end{figure}

\begin{figure}[t!]
\centering
\includegraphics[width=1\textwidth]{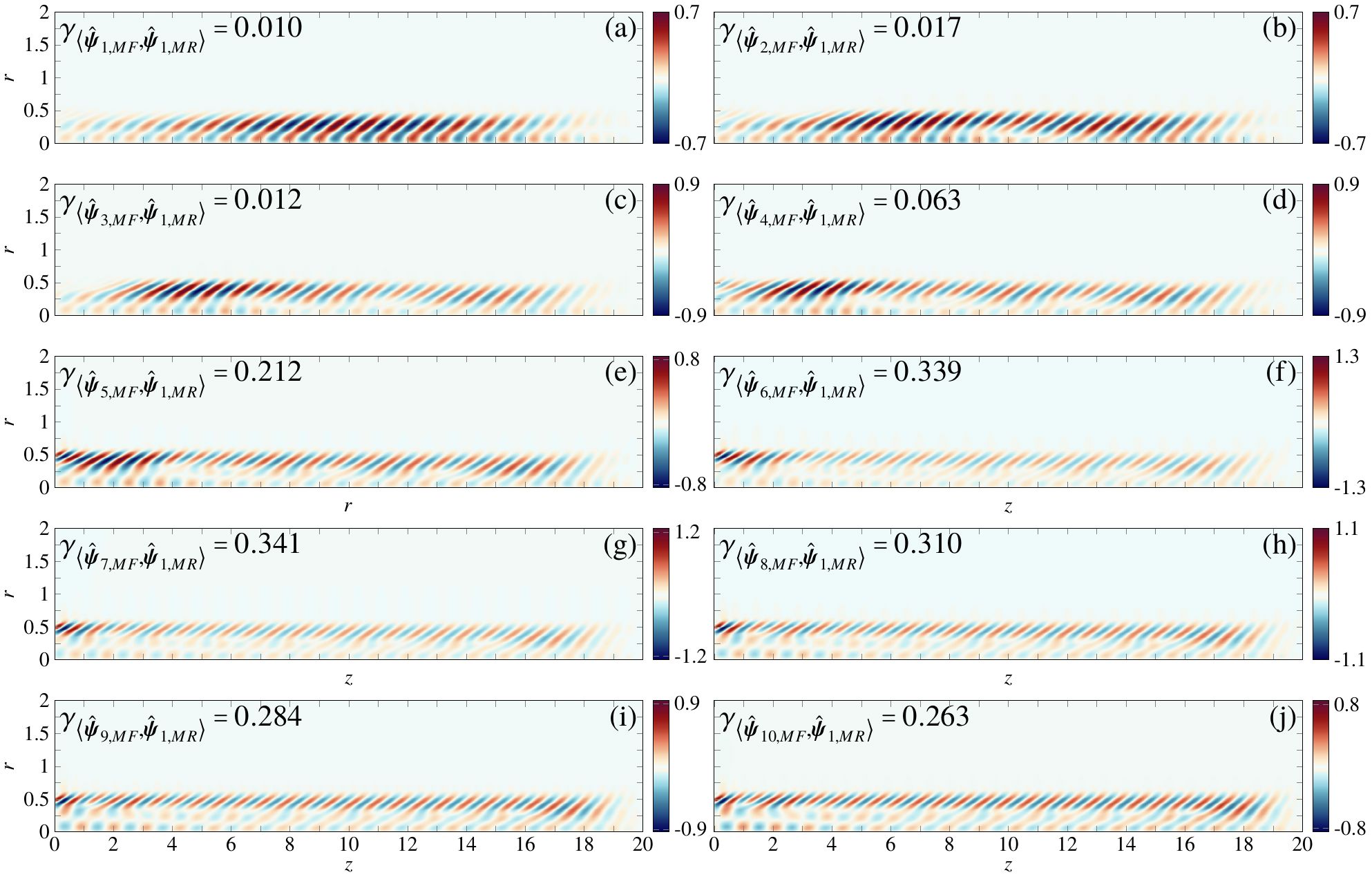}
\caption{Real part of the axial velocity component associated with the first ten suboptimal input modes $\hat{\boldsymbol{\psi}}_{j,MF}$ from mean-flow resolvent analysis, for the weakly unsteady case. These modes were computed for a forcing frequency $\omega=3\omega_0/2$, i.e. at the second subharmonic peak, as in Fig.~\ref{fig:Fig11}. Each mode has been normalised such that its phase at a coordinate $\left(r,z\right)=\left(0.5,0.1\right)$ is zero. The corresponding singular values are those shown in panel (c) of Fig.~\ref{fig:Fig8} as black circles (mean flow). The alignment coefficients between the optimal mean resolvent mode $\hat{\boldsymbol{\psi}}_{1,MR}$ and each suboptimal mean-flow mode $\hat{\boldsymbol{\psi}}_{j,MF}$ are also provided.}
\label{fig:Fig18} 
\end{figure}
 
As in the previous section, we start by examining the weakly unsteady case. The mean resolvent leading singular value obtained \textit{via} the projection method is shown in Fig.~\ref{fig:Fig8}(a) for different forcing frequencies (empty blue triangles). For a subspace dimension of $d=60$, we observe an excellent agreement with the results from the reference framework (solid red curve). This is not surprising at frequencies $\omega\gtrsim2\omega_0$ where the optimal gain curves from the two resolvent operators are nearly superimposed. But interestingly, convergence is fast even in cases where the optimal gains curves of the two operators are well separated. In the following, we will therefore comment mainly on the results at the sub-harmonic peak $\omega=\omega_0/2$ and its harmonic $\omega=3\omega_0/2$, where the largest discrepancies between $\mathsfbi{R}_{\overline{\boldsymbol{\mathrm{Q}}}}$ and $\mathsfbi{R}_0$ are found.\\
\indent At $\omega_0/2$, the high sensitivity of the system to sub-harmonic disturbances makes it low-rank; this is captured by both analyses, even though the optimal gain from $\mathsfbi{R}_{\overline{\boldsymbol{\mathrm{Q}}}}$, which overlooks interactions with the unsteady part of the base flow, is much lower than that from $\mathsfbi{R}_0$. Despite some qualitative differences, Fig.~\ref{fig:Fig6} showed that approximately the same spatial support characterises the optimal forcing mode from the mean resolvent and mean flow analysis; thus, using a subspace of mean flow input mode to build the vector basis and perform the projection seems indeed a suitable choice. In Fig.~\ref{fig:Fig9}, we display the matrix of projection coefficients $\boldsymbol{\Gamma}=(\gamma_{ij})$,  where we recall that each column $\boldsymbol{\gamma}_j$ is a unit-norm eigenvector of~\eqref{eq:smallghevp}. This matrix has a clear diagonal structure for a subspace dimension $d\lesssim 30$, meaning that optimal forcing mean-flow and mean resolvent modes are relatively well aligned (we recall the value $>0.85$ of the alignment coefficients reported in table~\ref{tab:Tab1}); for $d>30$ the projection starts to deteriorate and spread over the off-diagonal coefficients.\\
\indent \, Interestingly, at $\omega=\omega_0/2$, a single-mode projection ($d=1$), i.e. $\hat{\boldsymbol{\psi}}^1_{MR,1}=\hat{\boldsymbol{\psi}}_{MF,1}$, is already sufficient to significantly improve the gain prediction associated with the optimal response mode (see light-blue crosses in Fig.~\ref{fig:Fig8}(a) and the first gain value for $d=1$ in Fig.~\ref{fig:Fig10}(a)). This is explainable by noticing that, even though the input is the leading forcing mode of mean-flow resolvent analysis, interactions with the unsteady part of the flow are inherently captured by the mean resolvent operator (see Neumann expansion (\ref{eq:Neumann})) and intrinsically manifest in the database of computed responses $\hat{\mathsfbi{Y}}$ used to construct the reduced problem~\eqref{eq:smallghevp}.\\
\indent \, At $3\omega_0/2$, the mean-flow resolvent completely overlooks the secondary receptivity peak. The mean-flow resolvent does not capture either the large spectral gap of the mean resolvent between the optimal gain and the first suboptimal gain visible in Fig.~\ref{fig:Fig8}(c). These features are footprints of the base flow unsteadiness that mean-flow analysis cannot capture, and one may therefore anticipate that the projection method will fail at that frequency. Yet, the approximation already converges for $d\approx10$ (see blue markers in Fig.~\ref{fig:Fig8}(a)).\\
\indent \, The corresponding matrix of projection coefficients is shown in Figs.~\ref{fig:Fig9}(b) and~\ref{fig:Fig10}(b). The matrix does not have a diagonal structure as in panel (a) for $\omega_0/2$. In particular, a large number of mean-flow modes are necessary to capture the leading mean-resolvent modes (see first column of projection coefficients). The first four mean-flow modes appear to be nearly orthogonal to the mean resolvent mode. The same behaviour is observed when looking at the convergence of $\lambda^d_{1,MR}/\lambda_{1,MR}$ versus $d$ in Fig.~\ref{fig:Fig10}(b): the ratio only starts ramping up from 0 for $d\geq 5$ and then quickly converges to 1. These observations are consistent with the evolution of $\hat{\boldsymbol{\psi}}^d_{1,MR}$ versus $d$ in Fig.~\ref{fig:Fig11}. For $d=1$ (panel (a) of Fig.~\ref{fig:Fig11}), $\hat{\boldsymbol{\psi}}^d_{1,MR}=\hat{\boldsymbol{\psi}}_{1,MF}$, but we already noted in Fig.~\ref{fig:Fig6} that the spatial support of $\hat{\boldsymbol{\psi}}_{1,MF}$ (panel (b) therein) is quite different from that of $\hat{\boldsymbol{\psi}}_{1,MR}$ (panel (d) therein), inducing a poor alignment. From $d=5$ (panel (b) of Fig.~\ref{fig:Fig11}), the spatial support of the aproximation $\hat{\boldsymbol{\psi}}^d_{1,MR}$ starts shifting upstream and for $d=10$ (panel (c) of Fig.~\ref{fig:Fig11}), the mode is almost fully localized near the inlet. This behaviour can be understood by examining the spatial support of the mean-flow forcing modes $\hat{\boldsymbol{\psi}}_{j,MF}$ in Fig.~\ref{fig:Fig18}: these only start having a significant contribution close to the inlet for $j\geq 5$.\\  
\indent \, In Fig.~\ref{fig:Fig8}, we examine the 10 suboptimal gains at $\omega_0/2$ (panel (b)) and $3\omega_0/2$ (panel (c)), for the mean-flow resolvent (black circles), the mean resolvent (red squares) and the approximation with $d=60$ (blue triangles). We note that for that subspace dimension, all ten suboptimal gains of the mean resolvent are converged for both frequencies.\\

\begin{centering}\subsection{Strongly unsteady base flow}\label{subsec:SSSS2}\end{centering}

\begin{figure}[t!]
\centering
\includegraphics[width=0.85\textwidth]{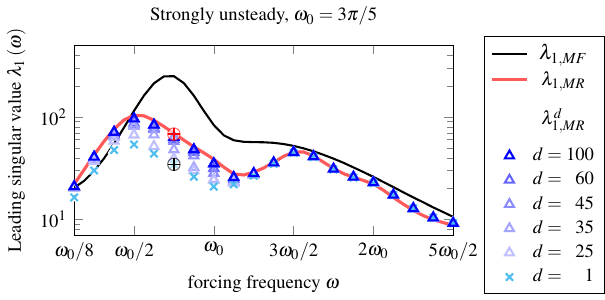}
\caption{Leading singular value $\lambda_1$, as a function of the forcing frequency $\omega$, for the strongly unsteady base flow. Black line: resolvent analysis about the mean flow. Red line: mean resolvent analysis computed for $\mathrm{N}_{\mathrm{h}}=5$. Blue empty triangles: optimal gain predicted by the projection method discussed in \S\ref{eq:mainsec} using increasing subspace dimensions up to $d=100$, optimal forcing mode from the resolvent analysis about the mean flow. The special case of a rank-one projection is indicated by the light-blue crosses. The circled plus signs correspond to mean resolvent gains $\lambda_{1,MR}$ and $\lambda^1_{1,MR}$ computed from time-marching the linearised equations forced harmonically by $\hat{\boldsymbol{\psi}}_{1,MR}$ (red) and $\hat{\boldsymbol{\psi}}_{1,MF}$ (black) at $\omega=3\omega_0/4$ (see \S\ref{subsubsec:AppAsub1}).}
\label{fig:Fig13bis} 
\end{figure}

The analysis is repeated for the strongly unsteady base flow, and the convergence of the leading mean resolvent gain is reported in Fig.~\ref{fig:Fig13bis}. Results are contrasted. For $\omega\geq 5\omega_0/4$, a single input mode appears sufficient to collapse the approximation with the red curve of mean resolvent analysis. But for $\omega<\omega_0/4$, convergence is very slow and the approximation becomes quantitative only for $d\geq 100$. However, it is interesting to note that even for $d=1$, the resonance peak at $\omega_0/2$ is correctly captured. Therefore, simply applying $\mathsfbi{R}_0$ to the leading forcing mode of the mean-flow resolvent $\hat{\boldsymbol{\psi}}_{1,MF}$ is sufficient to correctly capture the dominant receptivity peak associated with vortex pairing, which the mean-flow resolvent alone fails to identify. It is not at all surprising that convergence is slower for the strongly unsteady case, but it is interesting to note that there is already a qualitative benefit starting from $d=1$. Besides, $d=100$ remains in any case much smaller than the full input space dimension $\mathrm{N}=240,000$.\\

\begin{centering}\subsubsection{Consistency between matrix-based and matrix-free implementations}\label{subsubsec:AppAsub1}\end{centering}

\begin{figure}[t!]
\centering
\includegraphics[width=0.85\textwidth]{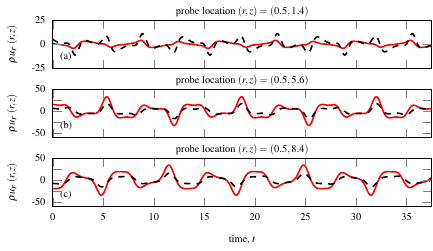}
\caption{Time series of the statistically-steady linear response probed at three different locations. Black dashed line: linear response to $\hat{\boldsymbol{\psi}}_{1,MF}\mathrm{e}^{\mathrm{i}\omega t}+\mathrm{c.c.}$. Red line: linear response to $\hat{\boldsymbol{\psi}}_{1,MR}\mathrm{e}^{\mathrm{i}\omega t}+\mathrm{c.c.}$.}
\label{fig:Fig15} 
\centering
\bigskip
\includegraphics[width=1\textwidth]{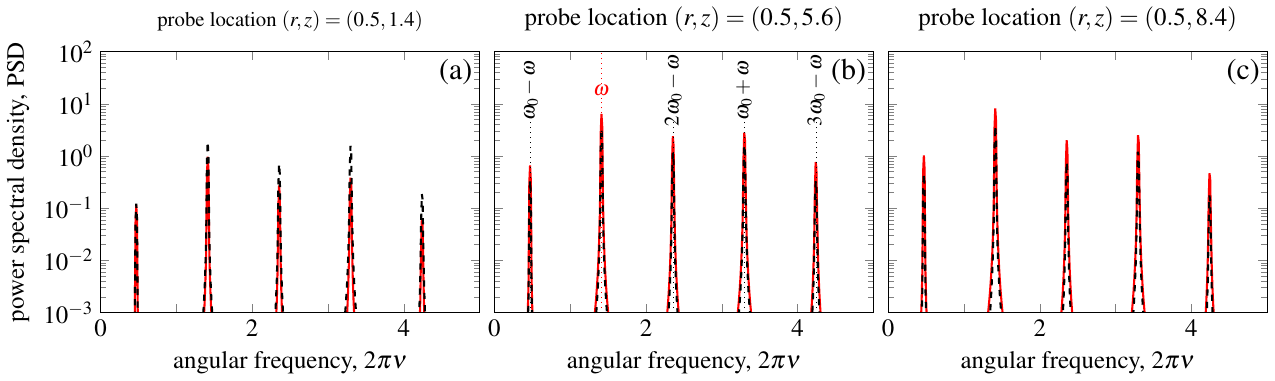}
\caption{Power Spectral Density (PSD) associated with the three probes of Fig.~\ref{fig:Fig15} and computed using a single DFT with Hann windowing. Linear responses to harmonic forcings by the optimal mean-flow resolvent input mode (black dashed) and optimal mean resolvent input mode (solid red).} 
\label{fig:Fig16} 
\end{figure}

So far, we have fully relied on matrix-based implementations of the adjoint-based and \textcolor{black}{projection-based} approaches. However, both approaches may also be implemented in a matrix-free, time-stepper fashion. In the present section, we simply illustrate the consistency between the two paradigms, by computing the linear response to harmonic forcing by the optimal forcing modes $\hat{\boldsymbol{\psi}}_{1,MF}$ and $\hat{\boldsymbol{\psi}}_{1,MR}$ of the two resolvent operators. We select a forcing frequency $\omega=3\omega_0/4$ approximately corresponding to the maximum gain of the mean-flow resolvent curve in Fig.~\ref{fig:Fig13bis}.\\
\indent \, Figure \ref{fig:Fig15} shows the timeseries of the statistically-steady linear responses for three probes at $r=0.5$ and $z=1.4,5.6,8.4$. The dashed curves correspond to forcing by the leading mean-flow resolvent mode, while the solid red curves correspond to forcing by the leading mean resolvent mode. We observe that the responses are not simply periodic. The power spectral densities of the various signals are evaluated using DFT on the full timeseries, using a single Hann window. The results are reported in Fig.~\ref{fig:Fig16}. As expected, the responses have energy at frequencies $\omega+n\omega_0$ for all integers $n$: the permanent response of the LTP system to harmonic forcing is an EMP signal, as indicated in \eqref{eq:EMPresp}. For the most upstream probe (a), the response to $\hat{\boldsymbol{\psi}}_{1,MF}$ is more energetic than the response to $\hat{\boldsymbol{\psi}}_{1,MR}$, but the opposite is true for downstream probes (b,c).\\
\indent \, Following the procedure described in \S~\ref{subsec:timedomain}, the component at the forcing frequency $\omega$ is extracted from the responses using harmonic averaging. In practice, an on-the-fly harmonic average is implemented in order to avoid the storage of $\boldsymbol{\mathrm{y}}'\left(t\right)$. As already explained, this amounts to computing the response of the mean resolvent operator to the harmonic forcing, yielding $\lambda_{1,MR}\hat{\boldsymbol{\phi}}_{1,MR}=\mathsfbi{R}_0\hat{\boldsymbol{\psi}}_{1,MR}$ and $\lambda^1_{1,MR}\hat{\boldsymbol{\phi}}^1_{1,MR}=\mathsfbi{R}_0\hat{\boldsymbol{\psi}}_{1,MF}$. Since $\hat{\boldsymbol{\phi}}_{1,MR}$ and $\hat{\boldsymbol{\phi}}^1_{1,MR}$ both have unit ($\mathsfbi{M}$-)norm, the output norms are respectively equal to $\lambda_{1,MR}$ and $\lambda^1_{1,MR}$. This is verified in Fig.~\ref{fig:Fig13bis}, where the circled plus signs indicate the two output norms at $\omega=3\omega_0/4$, following the aforementioned procedure. The red symbol denotes the response to forcing by $\hat{\boldsymbol{\psi}}_{1,MR}$ and is superimposed on the red curve previously computed with the matrix-based implementation of the adjoint-based approach. The black symbol denotes the response to forcing by $\hat{\boldsymbol{\psi}}_{1,MF}$ and is superimposed on the light blue cross marker previously computed with the matrix-based implementation of the \textcolor{black}{projection-based} approach.\\
\indent \, Using the time-stepper framework, the user has the choice to either implement an adjoint-based approach similar to that of Ref.~\onlinecite{farghadan2024efficient}, or the present projection-based approach. Comparing the computational cost of the two time-based approaches is out of the scope of the present study. The projection-based approach is obviously convenient if one is unable to time-step the adjoint linearised equations. In the weakly unsteady case, we saw that a subspace dimension of $d=10$ was already sufficient to reach quantitative results, and the 10 simulations may be run in parallel for any given frequency $\omega$. In contrast, adjoint-looping is sequential in nature, although time-stepping of the adjoint equations may be performed in parallel \citep{costanzo2022parallel}. However, the projection-based approach becomes less attractive in strongly unsteady cases where the subspace dimension may need to be potentially greater than 100 for accurate results, unless efficient parallelisation schemes can be implemented (this is outside the scope of the present paper).\\


\begin{centering}\section{Conclusions}\label{sec:CONCL}\end{centering}

This work proposes numerical methods to perform mean resolvent analysis of periodic flows. The goal is to obtain a low-rank representation of the operator, capturing optimal input-output mechanisms in a linear time-invariant framework. \textcolor{black}{We first show how this can be done using the harmonic resolvent framework. While the harmonic resolvent operator is specific to the periodic case, the mean resolvent operator is not. Therefore, we also propose a projection method which does not hinge on the harmonic resolvent framework, but on the mean-flow resolvent instead. We leverage the fact that the mean-flow resolvent approximates the mean resolvent in order to propose a projection approach. This approach does not require computing the action of the adjoint about the unsteady attractor, which may be convenient for future extensions to chaotic and turbulent cases. However, the present paper only deals with periodic flows and aims at assessing the convergence of the projection approach against the approach based on the harmonic resolvent.} \\
\indent Beyond the algorithmic goal explained in the first paragraph, the study is also strongly motivated by a physical objective: comparing the results of mean-flow resolvent and mean resolvent analyses on an open-shear flow, a task which has never been done before (although mean linear responses have been compared to linear responses about the mean flow in the time domain; see Refs.~\onlinecite{luchini2006phase,russo2016linear}). The mean-flow resolvent operator is rarely used in the context of flow control relevant to the present paper, as it arises much more often in the context of modelling of second-order statistics of turbulent, \textit{unforced} flows (see introduction for references). However, it is also the most straightforward option for modelling input-output behaviour of unsteady flows for control purposes. The mean resolvent operator is, by definition, the statistically optimal LTI operator for input-output analysis of statistically-steady flows; again, in the context of flow control where the forcing is exogenous (rather than endogenous as in McKeon \& Sharma \citep{mckeon2010critical}'s framework). So on the one hand, the mean-flow resolvent is obviously the go-to option from an engineering perspective because it is easy to obtain, but it suffers from intrinsic limitations recalled in the introduction. On the other hand, the mean resolvent is much trickier to compute, but optimal for our specific purpose: we were therefore motivated to quantify the gap between predictions from the two operators.\\
\indent \, The present paper addresses these two goals on the case of a nearly incompressible axisymmetric time-periodic laminar jet that was already studied in the context of harmonic resolvent analysis \citep{padovan2022analysis}, allowing for validation. The periodic base flow is not self-sustained but caused by periodic forcing at the inlet. We consider two forcing frequencies $\omega_0$ (not to be confused with $\omega$, the frequency of the \textit{linear} forcing) leading to qualitatively different base flows: one `weakly unsteady' in the sense that the Fourier harmonics of the base flow decay quickly, and a `strongly unsteady' case where the decay is slow. Using the reference method based on the harmonic resolvent operator, we find for the weakly unsteady case that both the mean-flow resolvent and the mean resolvent capture a dominant receptivity peak at $\omega_0/2$, a physically-relevant frequency related to vortex pairing. However, the associated gain is much stronger in the mean resolvent case. The mean resolvent also captures a secondary receptivity peak at $3\omega_0/2$, which is absent from the mean-flow resolvent analysis, due to the lack of modelling of interactions between the linear perturbations and the unsteady part of the base flow. As expected, the differences between the two operators are more pronounced in the strongly unsteady case. In that case, the mean resolvent still displays a dominant gain peak at $\omega_0/2$ and a secondary peak at $3\omega_0/2$, which are physically meaningful. But the dominant receptivity frequency of the mean-flow resolvent seems physically irrelevant, as it is located at an intermediate frequency $\omega \approx 3\omega_0/4$. For the strongly unsteady case, the maximum of the mean resolvent is less than that of the mean-flow resolvent, while the opposite was true in the weakly unsteady case. The alignment coefficient between the optimal forcing modes of the two operators at $\omega_0/2$ and $3\omega_0/2$ is less than 0.5 for the strongly unsteady case, revealing a significant discrepancy in the modelling of optimal receptivity mechanisms. Surprisingly, though, the alignment of the corresponding optimal responses is higher than 0.95 in both cases.\\
\indent \, \textcolor{black}{The projection method was then validated on the results of the reference method.} To reach a correct estimation of the optimal gain, a subspace dimension of $d=10$ appears sufficient in the weakly unsteady case, whereas a dimension of $d=100$ is a minimum for the strongly unsteady case. This is consistent with the poor alignment between the leading optimal forcing modes of the two operators in the strongly unsteady case. However, despite this discrepancy, it was noted that the dominant receptivity mechanisms at $\omega_0/2$ and $3\omega_0/2$ can already be captured in the gain curve with an input subspace of dimension 1, i.e. considering the mean response to the optimal forcing of the mean-flow resolvent alone. So, at a very low cost, the method is already able to bring a significant correction to the mean-flow resolvent analysis for our purposes.\\
\indent \, Future work may consider ways to learn better projection bases for faster convergence of the approximation in strongly unsteady cases. More importantly, the present \textcolor{black}{projection-based time-stepping} approach needs to be extended to more general stochastic and chaotic frameworks applicable to turbulent flows. It will also be interesting to investigate whether the mean resolvent operator has any relevance for predicting second-order statistics of unforced flows, in connection with much of the ongoing work with the mean-flow resolvent and eddy-viscosity calibration.\\


\bigskip
\begin{centering}\section*{ACKNOWLEDGMENTS}\end{centering}
\indent The authors acknowledge Dr Samir Beneddine for the fruitful insights about the numerical implementation of the parallelised GMRES solver in PETSc/SLEPc. The authors also acknowledge Dr Javier Sierra-Aus\'in for stimulating discussions on linear response theory.\\

\begin{centering}\section*{AUTHOR DECLARATIONS}\end{centering}
\begin{centering}\subsection*{Conflict of Interest}\end{centering}
\indent The authors report no conflict of interest.\\

\begin{centering}\section*{DATA AVAILABILITY}\end{centering}
\indent The data that support the findings of this study are available from the corresponding author upon reasonable request.\\


\appendix

\begin{centering}\section{Validation of the numerical tools}\label{sec:AppB}\end{centering}

The present Appendix is devoted to the validation of the numerical tools employed in this study. This is done considering the weakly unsteady jet configuration oscillating at $\omega_0=6\pi/5$, for which a direct comparison with Ref.~\onlinecite{padovan2022analysis} (P\&R) is possible.\\

\begin{centering}\subsection{Grid convergence for mean-flow analysis}\label{subsec:Sec3sub4_0}\end{centering}

\begin{figure}[t!]
\centering
\includegraphics[width=1\textwidth]{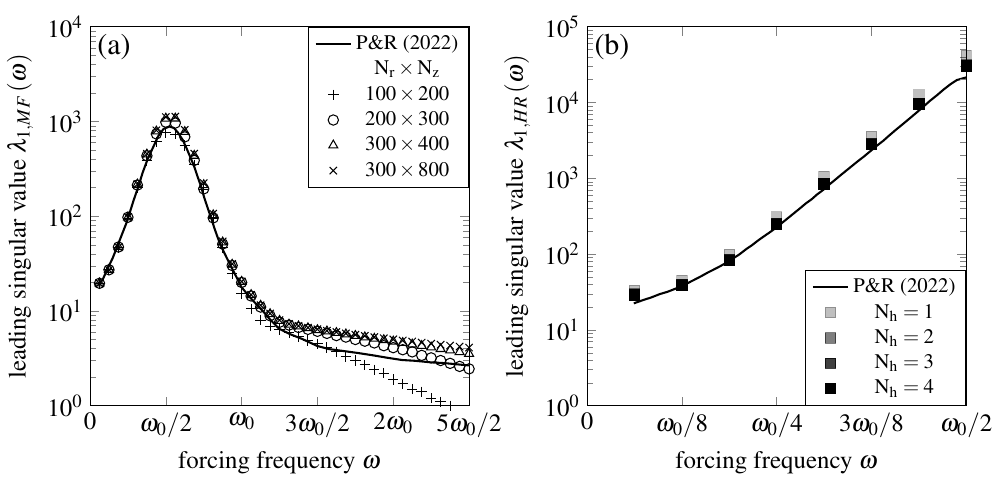}
\caption{(a) Leading singular value $\lambda_{1,MF}\left(\omega\right)$ predicted by the resolvent analysis about the mean flow for the axisymmetric jet forced by an axial inflow velocity oscillating periodically described at $\omega_0=6\pi/5$ (\S\ref{sec:Sec3}). The Reynolds number is $Re=1000$. The analysis is performed on an interval of forcing frequency $\omega$ ranging from $0.1$ to $5/2$ times the base flow frequency. The black solid line corresponds to the mean flow analysis of Ref.~\onlinecite{padovan2022analysis} (P\&R). Markers correspond instead to the present results obtained for four different grid refinements. The finest mesh is the same as P\&R. (b) Harmonic resolvent leading singular value $\lambda_{1,HR}\left(\omega\right)$ as computed by P\&R for $Re=1000$ (black solid line) and from the present analysis (markers) for $\mathrm{N}_{\mathrm{h}}=1,2,3$ and $4$ over the relevant set of frequencies $0\le\omega\le\omega_0/2$, with $\omega_0=6\pi/5$. A good convergence is achieved already for $\mathrm{N}_{\mathrm{h}}=2$ (the data for $\mathrm{N}_{\mathrm{h}}=3$ and $4$ are indistinguishable from those for $\mathrm{N}_{\mathrm{h}}=2$). For this computation, a mesh of size $200\times 300$ was used.}
\label{fig:Fig4} 
\end{figure}

\indent Concerning the mean-flow analysis, in Fig.~\ref{fig:Fig4}(a), we report the leading singular value $\lambda_{1,MF}$ as a function of the forcing frequency $\omega$ and computed for four different grids as explained in~\S\ref{subsec:linalg}. Results are robust and agree fairly well with the reference case from P\&R (black solid line). The dependence on the mesh refinement is mostly visible at higher frequencies, with a grid $\mathrm{N}_{\mathrm{r}}\times \mathrm{N}_{\mathrm{z}}=200\times 300$ ($\mathrm{N}=4\times \mathrm{N}_{\mathrm{r}}\times \mathrm{N}_{\mathrm{z}}=240\,000$) that provides a satisfactory trade-off. We notice how, even for the finest mesh, the present results do not perfectly overlap with the reference case; such discrepancy could be caused by the weak compressibility effects allowed for in our formulation at $M=0.1$, although a more likely explanation could reside in the fact that the authors of Ref.~\onlinecite{padovan2022analysis} used a third-order upwind scheme to discretise the advective term, whereas we used a fifth-order scheme (see the comparison between Refs.~\onlinecite{bugeat20193d} and~\onlinecite{poulain2023broadcast} on the strong effect of the scheme order on the convergence of results). Nevertheless, the discrepancy is mostly visible at frequencies characterised by a relatively low gain when compared to the predicted dominant peak found at $\omega\approx\omega_0/2$.\\

\begin{centering}\subsection{GMRES solver with block-Jacobi preconditioner for harmonic resolvent analysis}\label{subsec:Sec3sub4}\end{centering} 

As commented in~\S\ref{subsec:linalg}, in this study, mean resolvent analysis has been computed in the frequency domain using an iterative solver based on GMRES algorithms \citep{rigas2021nonlinear,poulain2024adjoint} with a block-Jacobi preconditioner, which is efficient for diagonally-dominant matrices, i.e. for weakly unsteady flows. The code was also made parallel and distributed using PETSc, with each processor core handling a $\left(\mathrm{M}\times \mathrm{M}\right)$-block, corresponding to a single Fourier component, for which a sparse direct LU method was applied using the external package MUMPS \citep{amestoy2000mumps}.\\
\indent \, To validate our numerical implementation against Ref.~\onlinecite{padovan2022analysis}, we perform harmonic resolvent analysis. To do so, we consider EMP inputs and outputs described as vectors of frequency component $\hat{\boldsymbol{\mathcal{F}}}$ and $\hat{\boldsymbol{\mathcal{Y}}}$ in \eqref{eq:eq13}, and choose $s=\mathrm{i}\omega$. We maximise the energy gain $\|\hat{\boldsymbol{\mathsfbi{\mathcal{Y}}}}\|^2/\|\hat{\boldsymbol{\mathsfbi{\mathcal{F}}}}\|^2$ over the set of non-zero input vectors $\hat{\boldsymbol{\mathcal{F}}}$. The $\boldsymbol{\mathcal{M}}_f$-orthogonal basis of optimal forcing modes solves the following very large generalised eigenvalue problem
\begin{equation}
\label{eq:eq10}
\left(\boldsymbol{\mathsfbi{\mathcal{R}}}^H\boldsymbol{\mathsfbi{\mathcal{M}}}_y\boldsymbol{\mathsfbi{\mathcal{R}}}\right) \hat{\boldsymbol{\psi}}_{j,HR} = \lambda_{j,HR}^2 \boldsymbol{\mathsfbi{\mathcal{M}}}_f \hat{\boldsymbol{\psi}}_{j,HR},
\end{equation}
where $\lambda_{j,HR}^2$ are the corresponding energy gains. Here $\boldsymbol{\mathsfbi{\mathcal{M}}}_y$ and $\boldsymbol{\mathsfbi{\mathcal{M}}}_f$ are block-diagonal matrices formed from $\mathsfbi{M}_y\in\mathbb{R}^{\mathrm{K}\times\mathrm{K}}$ and $\mathsfbi{M}_f\in\mathbb{R}^{\mathrm{M}\times\mathrm{M}}$, respectively. The associated response is obtained as $\hat{\boldsymbol{\phi}}_{j,HR}=\boldsymbol{\mathsfbi{\mathcal{R}}}\hat{\boldsymbol{\psi}}_{j,HR}/\lambda_{j,HR}$.\\
\indent \, Results are reported in Fig.~\ref{fig:Fig4}(b). The number of harmonics retained is progressively increased from $\mathrm{N}_{\mathrm{h}}=1$ to $\mathrm{N}_{\mathrm{h}}=4$. We observe that $\mathrm{N}_{\mathrm{h}}=2$ is already sufficient to ensure a fair convergence and a good agreement with the reference case from Ref.~\onlinecite{padovan2022analysis}. Note that harmonic resolvent analysis is fully determined over the set of forcing frequencies $0\le \omega < \omega_0$ (see Ref.~\onlinecite{padovan2022analysis}).


\bibliographystyle{unsrt}
\bibliography{Bibliography}


\end{document}